\documentclass[10pt,aps,prl,twocolumn,amsmath,amssymb,floatfix,superscriptaddress]{revtex4-2}

\usepackage{caption}
\usepackage{graphicx}
\usepackage{subcaption}
\usepackage{physics}
\usepackage[normalem]{ulem}
\usepackage[title]{appendix}
\usepackage{hyperref}
\usepackage{comment}

\usepackage[usenames,dvipsnames]{color}

\newcommand{\affA}{Van der Waals-Zeeman Institute, Institute of Physics, University of Amsterdam, 1098 XH Amsterdam, Netherlands}
\newcommand{\affB}{QuSoft, Science Park 123, 1098 XG Amsterdam, the Netherlands}

\newcommand{\affD}{Institute for Quantum Electronics, ETH Z\"urich, Otto-Stern-Weg 1, 8093 Z\"urich, Switzerland}
\newcommand{\affE}{Quantum Center, ETH Z{\"u}rich, 8093 Z{\"u}rich, Switzerland}

\begin{document}

\title[Sample title]{Quadratic spin-phonon coupling and bipolarons in trapped ions}
% Force line breaks with \\
\author{L.~P.~H. Gallagher}\affiliation{\affA}
\author{M. Mazzanti}\affiliation{\affD}\affiliation{\affE}
\author{Z.~E.~D.~Ackerman}\affiliation{\affA}
\author{R.~J.~C. Spreeuw}\affiliation{\affA}
\author{A. Safavi-Naini}\affiliation{\affA}\affiliation{\affB}
\author{R. Gerritsma}\affiliation{\affA}\affiliation{\affB}
%\email{r.x.schuessler@uva.nl}

\begin{abstract}
We consider the quantum simulation of quadratic spin-phonon coupling in a crystal of trapped ions. The coupling is implemented using tightly focused optical tweezers on each ion that change the local trapping potential in a state-dependent way. By encoding spins in the internal states of the ions and adding a tunneling term via M{\o}lmer-S{\o}rensen-type interactions, we calculate the emergence of mobile bipolarons driven by the zero-point energy of the ion crystal phonons. We show that thermal occupation may pin the bipolarons for ion crystals at finite temperature. Our scheme can be used to study and illustrate the emergence of mobile bipolarons as a function of temperature.
\end{abstract}

\date{\today}
\maketitle

%\section{Introduction}
%\label{sec:level1}

\emph{Introduction}.---Trapped ions provide precise control over their quantum state, and have emerged as one of the leading platforms for the study of quantum phenomena. Laser-induced state-dependent forces on the ions are used to excite collective vibrations in the ion crystal. These in turn give rise to effective long-range  interactions between the internal states of the ions. Using two or more internal states to encode (pseudo-)spin makes it possible to perform quantum simulation of spin-models~\cite{Porras_2004,Friedenauer:2008,Kim:2009,Jurcevic:2014,Monroe_Correlations_2014,Zhang:2017, Bohnet:2016} and to construct quantum computing platforms~\cite{Postler:2022,Postler:2024,Egan:2021,Pogorelov:2021,Wright:2019,Gustiani:2024,Ballance:2016}. 

In typical setups, the laser-ion interactions are linearly proportional to the excursion from the ion's equilibrium positions.  These linear spin-phonon couplings and the resulting effective spin-spin interactions have been extensively studied in trapped ion systems~\cite{Molmer:1999,Porras_2004,Friedenauer:2008,Kim:2009,Jurcevic:2014,Monroe_Correlations_2014,Zhang:2017}. Recently, there has been increasing interest in studying quadratic spin-phonon coupling~\cite{Mazzanti:2021,Katz:2022,Katz:2023}. While this coupling can be exploited to expand the scope of trapped ion quantum simulators~\cite{Leibfried:2002,Mezzacapo:2012}, achieving laser-ion couplings that are proportional to the square of the excursion from the ion's equilibrium positions are comparatively harder to achieve, as it requires obtaining a sizeable potential curvature over the size of the ion's wavepacket~\cite{Leibfried:2003}. 

In this letter, we propose to use long-lived internal states of ions, in combination with state-dependent tweezer potentials~\cite{Grier:2003,Scholl:2021,Bluvstein:2021,Omran:2019,Ashkin:1986, Beugnon:2007,Barredo:2016, Endres:2016, Omran:2019, Graham:2019, Kaufman:2021,Stopp_ang_mom2022,Urech:2022,bluvstein:2024,Teoh:2021, Espinoza:2021, Nath:2015, Olsacher:2020, Bond:2022, Mazzanti:2021,Schwerdt:2024}, to generate quadratic coupling between the internal and motional states of an ion crystal. We show that the resulting system can be used to simulate a class of Bose-Hubbard models, with long-range tunneling and tunable onsite two-body interactions emerging from quadratic phonon coupling. The long-range tunneling is generated via a M{\o}lmer-S{\o}rensen-type scheme~\cite{Monroe_Correlations_2014,Jurcevic:2014}. We calculate the emergence of mobile bipolarons in an ion crystal that is cooled close to its ground state. Here the zero-point energy of the phonons is responsible for the occurence of mobile bipolarons. The mechanism is analogous to the case in Ref.~\cite{Han:2024}, in which Han et al. studied the emergence of superconductivity in systems with quadratic electron-phonon coupling. In contrast to the linear electron-phonon coupling, here the emergence of superconductivity was shown to be the result of the formation of bipolarons and the quadratic nature of the coupling resulted in a different scaling in terms of the critical temperature and the mass of the lattice ions~\cite{Han:2024,Zhang:2023,Zhang:2024,Ragni:2023}. Finally, we consider the role of thermal occupation of phonon modes and show that the bipolarons get pinned in analogy to recent quantum simulations of localization in spin models~\cite{Smith:2016,Morong:2024}. This pinning may be studied in an experiment by straightforward adjustment of the laser cooling parameters.

\begin{figure}[b]
    \centering
    \includegraphics[width=\linewidth]{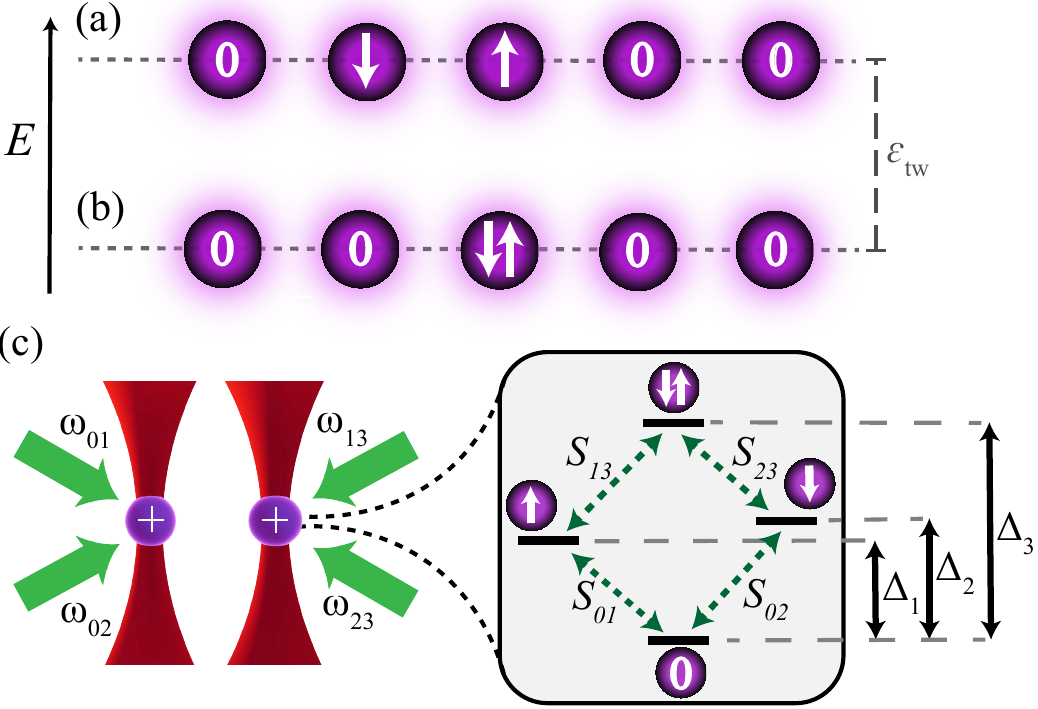}
    \caption{
    Spins are encoded in the internal states of the ions. Two ion state configurations are shown, with double occupation on a single site (b) energetically preferred compared to single occupation over two sites (a) by $\varepsilon_{\text{tw}}$, due to the tweezers. (c) Physical implementation: tweezers (in red) address the ions, and four global beams (in green) cause spin-spin interactions between the ions via a M{\o}lmer-S{\o}rensen-type scheme as described below. The laser detunings $\Delta_k$ are chosen such that only pairwise processes that conserve the encoded particle number and total spin are resonant as explained in the text.}
    \label{figure1}
\end{figure}

\emph{Tweezer setup}.---We consider a linear ion crystal with $N$ ions confined in a harmonic trap in combination with optical tweezers centered around the ion's equilibrium positions $Z_i^{(0)}$ to generate state-dependent potentials~\cite{Mazzanti:2023}. Each ion features four long-lived energy levels to encode the spin on each site. We identify the states as vacuum $\ket{0}$, single occupation $\ket{\uparrow}=\ket{1}$ and $\ket{\downarrow}=\ket{2}$, and double occupation $\ket{\uparrow\downarrow}=\ket{3}$, as shown in Fig. \ref{figure1}. In this mapping the four internal levels can be thought of as states in a two-component bosonic mixture with hardcore intraspecies and softcore interspecies interactions. This allows us to neglect the $\ket{\uparrow \uparrow}$ and $\ket{\downarrow \downarrow}$ states.

Next, we describe the emergence of quadratic spin-phonon. To simplify the discussion of the tweezers we only consider motion along the ion crystal (which we denote as the $z$-direction), but the process can be straightforwardly generalized to three dimensions.
The Hamiltonian of the system is given by: 
\begin{align}
    H=&H_\text{h}
    +\sum_{i}\left(\frac{\hat{p}_i^2}{2M}+\frac{1}{2}M\omega_z^2\hat{z}_i^2+\frac{1}{2}M\hat{O}_i\left(\hat{z}_i-Z_i^{(0)}\right)^2\right)\nonumber\\
    &+\frac{1}{2}\sum_{j\neq i}\frac{e^2}{4\pi\epsilon_0\vert  z_i-z_j\vert}\label{eq_Hgeneral}.
\end{align}
Here $H_\text{h}$ describes hopping of spin excitations from one ion to another, as we describe below. The position and momentum operators of ion $i$ are denoted by $\hat{z}_i$ and $\hat{p}_i$ respectively. The first summation describes the harmonic Paul trap and the tweezer potentials, with $\omega_z$ the trap frequency and $M$ the mass of one ion. The operator $\hat{O}_i$ describes the squared tweezer trap frequency for each ion $i$ and its dependence on the internal state. 
The second summation is the Coulomb interaction between the ions, with $e$ the elementary charge, and $\epsilon_0$ the vacuum permittivity. 

We can obtain the equilibrium positions for each ion $ Z_i^{(0)}$ by solving $\nabla V=0$~\cite{James:1998}, with $V$ the total potential energy. Each of the tweezers is centered on the equilibrium position of an ion, and has negligible effect on the neighbouring ions. Thus, the equilibrium positions do not depend on the tweezer potentials. In the absence of tweezers, the eigenmodes of the collective motion can be obtained by diagonalising the Hessian matrix, $D_{ij}=(1/M)d^2V/(dz_i dz_j)$ evaluated at the equilibrium positions $Z_i^{(0)}$~\cite{James:1998}. The eigenvalues $\lambda_m$ of the Hessian correspond to the squared vibrational frequencies $\omega_m=\sqrt{\lambda_m}$ in the absence of the tweezers, while the normalised eigenvectors with elements $b_{mi}$ describe the relative amplitude of motion of ion $i$ in mode $m$, $\hat{z}_i=\sum_m b_{mi}\hat{z}_m$. Here, we have also made the replacement $\hat{z}_i\rightarrow \hat{z}_i-Z_i^{(0)}$ such that the coordinates of each ion are defined with respect to their equilibrium position. In these coordinates, and to second order in ion position, the Hamiltonian is given by:
\begin{equation}
    H=H_\text{h}
    +\sum_{i}\left(\frac{\hat{p}_i^2}{2M}+\frac{1}{2}M\left(\omega_z^2+\hat{O}_{i}\right)\hat{z}_i^2+\sum_j V_{ij}\right).\nonumber
\end{equation}
The term $V_{ij}\propto \hat{z}_i\hat{z}_j$ describes interactions between the oscillators which are derived from the Coulomb interaction term.

Next we consider the effect of the state-dependent tweezer potential. For the case where the tweezer potential is weak compared to the ion trap, we can use first order perturbation theory to compute the new phonon frequencies, given by:
$\hat{\omega}_m\approx \sqrt{\omega_m^2+\sum_i \hat{O}_i b_{mi}^2 }$. Note that these frequencies are now also an operator on the internal states of the ions, i.e. the mode frequencies depend on the internal  states of all ions. %Comparing to Eq.~\ref{eq_Hgeneral}, we see that $\hat{O}_{i}=\sum_mb_{im}^2\hat{A}_i$.
Finally, we use the time-dependent unitary transformation $U(t)=e^{-i\sum_m\omega_m \hat{a}_m^{\dag}\hat{a}_m t}$ into a frame rotating with the unperturbed phonon frequencies $\omega_m$ and arrive at the Hamiltonian  due to the state-dependent tweezers, 
\begin{equation}
    H_\text{tw}=\sum_m \hbar(\hat{\omega}_m-\omega_m)\left(\hat{a}_m^{\dag}\hat{a}_m+1/2\right), \label{eq_Hwithphonons}
\end{equation} 
where $\hat{a}_m^{\dag}$ and $\hat{a}_m$ are the usual bosonic creation and annihilation operators of mode $m$. 

\emph{Bipolaron emergence}.---Bipolarons form when the $\ket{\uparrow \downarrow}$ state is energetically favourable compared to two singly occupied sites, and this energy is site independent, ensuring the mobility of the bipolaron. 
Since the zero point energy of the phonons drives the creation of mobile bipolarons we will focus on the phononic ground state in which we set $\hat{a}_m^{\dag}\hat{a}_m=0$ for all $m$. Before considering the form of $\hat{O}_i$, we first simplify the tweezer Hamiltonian in Eq.~\eqref{eq_Hwithphonons} further by Taylor expanding the square root, $\hat{\omega}_m\approx \omega_m+\frac{1}{2\omega_m}\sum_i \hat{O}_i b_{mi}^2$. The Hamiltonian then becomes $H\approx \sum_{m,i}\frac{1}{4\omega_m}\hat{O}_i b_{mi}^2$. We note that the expansion of the square root to first order neglects long-range interactions between the ion's internal states of order $\sim b_{mi}^2b_{mj}^2\hat{O}_i\hat{O}_j/(8\omega_m^3)$. However, we take these into account in the numerical analysis below. 

\begin{figure}[t]
    \centering
\includegraphics[width=\linewidth]{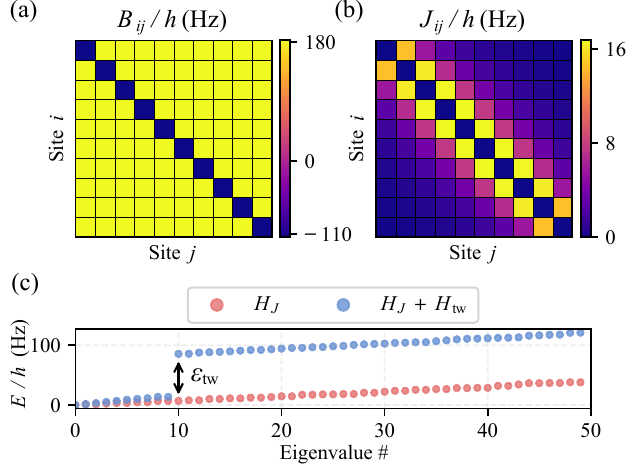}

\caption{(a) Tweezer matrix showing the ground state energy $B_{ij}$ of 2 opposite spins at sites \(i\) and \(j\). Spins paired at a site have minimum energy, and off-diagonal variations are under \(1\%\). Site homogeneity is achieved by varying tweezer intensities along the ion chain. The relative values are given as (0.61, 0.80, 0.91, 0.97, 1, 1, 0.97, 0.91, 0.80, 0.61). Note that for two spins, the diagonal components $B_{ii}\sim\gamma$, while the off-diagonal components correspond to $B_{i\neq j}\sim2g$ in equation~\ref{eq_Hgs}. (b) Tunneling matrix $J_{ij}$ between sites \(i\) and \(j\) showing predominantly nearest-neighbour tunneling. (c) First 50 eigenvalues of the Hamiltonian with the tweezer turned on \& off. The tweezer causes a gap between bipolaron states and the rest of the spectrum. See text for the parameters used.}
\label{energies}
\end{figure}

We consider the case in which $\hat{O}_i$ is a diagonal operator corresponding to a tweezer laser frequency that is tuned far from any atomic transition. Then the four elements on each site are proportional to the tweezer intensity~\cite{Grimm:2000} with $\hat{O}_i=\sum_{k=0}^3\varpi^2_{i,k}\text{sign}(\varpi_{i,k})|k\rangle_i\langle k|$, where $\varpi_i$ denotes the tweezer trap frequency for ion $i$ and state $|k\rangle$ and $|k\rangle_i\langle k|$ is the projector on state $k$ and site $i$. Here we use the negative sign to denote anti-confining tweezers. Using hollow tweezer potentials avoids homogeneous Stark shifts between the internal states of the ions~\cite{Stopp_ang_mom2022,Mazzanti:2021}.
We set the relative tweezer intensities on each site such that $\varpi^2_{i,3}=4\gamma/(\sum_m b_{mi}^2/\omega_m)$ which makes the Hamiltonian homogeneous to first order over the ion crystal, assuring spin mobility once we add a tunneling term. Setting $\varpi^2_{i,0}=0$ for simplicity and  $\varpi^2_{i,1}=\varpi^2_{i,2}=4g/(\sum_m b_{mi}^2/\omega_m)$. The Hamiltonian for the ground state (gs) is now given by:
\begin{equation}\label{eq_Hgs}
    H_{\rm gs}=\sum_{i=1}^Ng\Bigl(\ket{\downarrow}_i\bra{\downarrow}+\ket{\uparrow}_i\bra{\uparrow}\Bigr)+\gamma \ket{\uparrow\downarrow}_i\bra{\uparrow\downarrow}.
\end{equation}
This effective Hamiltonian will energetically favor a pair on one site over single occupation of two sites when $2g>\gamma$. In this case, the groundstate manifold is given by $N$ bipolaron states, separated by energy $\varepsilon_{\text{tw}}\sim 2g-\gamma$ from the rest of the spectrum, as shown in figure~\ref{energies}.

\emph{Hopping}.---We can implement the hopping term $H_\text{h}$ using a variation on the M{\o}lmer-S{\o}rensen scheme~\cite{Porras_2004,Friedenauer:2008,Kim:2009,Jurcevic:2014}. We couple the four levels in each ion using four laser fields with frequencies $\omega_{01}+\omega_{13}=\omega_{02}+\omega_{23}$, where the subscript denotes the levels that are coupled. The state-dependent force is generated by amplitude modulating the laser with frequency $\mu$. Within the interaction picture and after making the rotating wave approximation, the resulting Hamiltonian is given by, 
\begin{align}   H_\text{L}=&\sum_i\hbar\Omega k\hat{x}_i\cos(\mu t)\left(\hat{S}^{(i)}_{01}+\hat{S}^{(i)}_{02}+\hat{S}^{(i)}_{13}+\hat{S}^{(i)}_{23}+h.c.\right)\nonumber\\&+\Delta_{1}\ket{1}_i\bra{1}+\Delta_{2}\ket{2}_i\bra{2}+\Delta_{3}\ket{3}_i\bra{3}\label{Eq_hop}, 
\end{align}
where $\Omega$ denotes the Rabi frequency, which we set to be the same for each transition and all ions, $k$ is the effective wavenumber of the laser, and $\hat{x}_i$ the position of the ion with respect to its equilibrium position in the transverse direction. Here $\hat{S}^{(i)}_{jk}=|j\rangle_i\langle k|$ couples the states $j$ and $k$ of the $i$-th ion. The detunings are given by: $\Delta_1=E_1-\hbar\omega_{01}$, $\Delta_2=E_2-\hbar\omega_{02}$ and $\Delta_3=E_3-\hbar(\omega_{01}+\omega_{13})$, where $E_k$ denotes the energy of level $k$ and we set $E_0=0$ for simplicity. We assume that the levels are sufficiently separated in energy such that we can neglect the coupling of the laser fields to states other than the ones targeted.

Using $\hat{x}_i=\sum_m \beta_{mi}\hat{x}_m$, and following Ref.~\cite{Kim:2009}, the first term in Eq.~\ref{Eq_hop} leads to pairwise interactions of the form:
\begin{equation}\label{spinspin}
H_J=\sum_{ij}J_{ij}\hat{\mathbf{S}}_i\hat{\mathbf{S}}_j,
\end{equation}
\noindent with $\hat{\mathbf{S}}_i=\hat{S}^{(i)}_{01}+\hat{S}^{(i)}_{02}+\hat{S}^{(i)}_{13}+\hat{S}^{(i)}_{23}+h.c.$. The strength of each coupling is given by: 
\begin{equation}
    J_{ij}= \hbar\Omega^2\sum_m \frac{\beta_{mi}\beta_{mj}\eta^2_{m}\omega_m^{\perp}}{\mu^2-\lambda_m^{\perp}}.
\end{equation}
\noindent Here $\beta_{mi}$ denotes normalized radial eigenmode vectors, i.e. the relative amplitude of motion of ion $i$ in the radial mode $m$ with mode frequency $\omega_m^{\perp}=\sqrt{\lambda_m^{\perp}}$. The Lamb-Dicke parameters are given by $\eta_m=k\sqrt{\hbar/(2M\omega_m^{\perp})}$, with $k$ the effective wavenumber of the laser. 

Eq.~\ref{spinspin} describes pairwise flips of internal ion states. Out of the resulting 64 spin-spin interactions, only those that preserve the occupation number and total spin encoded across the ion chain are made resonant. This is achieved by setting the proper detunings $|\hbar\Delta_k|\gg |J_{ij}|$ for all $k$, $i$ and $j$. For instance, to preserve the total spin we choose $\Delta_1\neq\Delta_2\neq 0$, which ensures spin-flipping processes such as $\ket{10}\rightarrow\ket{02}$ are energetically suppressed. Likewise, processes that do not preserve particle number, such as $\ket{00}\rightarrow\ket{12}$, are suppressed by requiring $\Delta_1+\Delta_2=\Delta_3\neq 0$. Finally, setting $\Delta_k\ll |\mu-\omega_m|$ for all $k$ and $m$ assures that the effective Hamiltonian~(\ref{spinspin}) remains valid to lowest order in the presence of the laser detunings $\Delta_j$~\cite{Jurcevic:2014}.  

Our Hamiltonian allows hopping of spins to sites beyond nearest neighbor. The scaling with distance of the coupling may be approximated as $J_{ij}\propto 1/|i-j|^\alpha$, with $\alpha$ between 0 and 3, depending on the choice of $\mu$~\cite{Britton:2012,Monroe_Correlations_2014,Bohnet:2016}. In total, the effective Hamiltonian becomes: 
\begin{equation}\label{eq_Heff}
H_\text{eff}=\sum_{ijkk'}J_{ij}\left(\hat{S}^{(i)}_{k}\hat{S}^{(j)}_{k'}+h.c.\right)+H_\text{tw},
\end{equation}
\noindent  here we sum $\{k,k'\}=$ over all resonant pairwise processes, i.e. $\{k,k'\}=$\{(01,10),(02,20),(12,03),...\}. Note that bipolaron hopping ($\ket{30}\rightarrow \ket{03}$), while resonant, is not a pairwise interaction, but arises once we add the quadratic spin-phonon coupling, in analogy to the case of quadratic electron-phonon coupling in Ref.~\cite{Han:2024}.

\begin{figure*}[t]
    \centering
    \includegraphics[width=\linewidth]{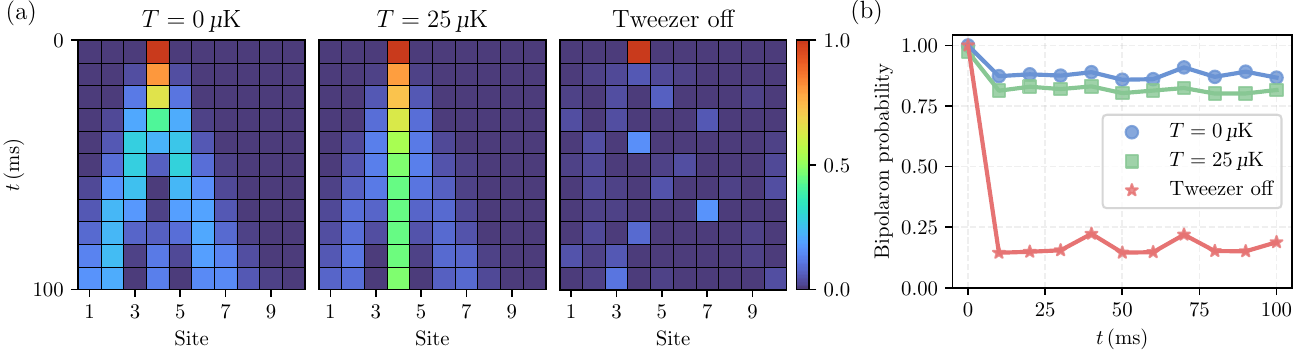}
    \caption{(a) Evolution of the bipolaron probability initially at site 4, with the ion crystal temperature \(T=0\)~K and \(T=25\)~$\mu$K (corresponding to \(\bar{n}_{\text{com}}=0.43\)), and without the tweezers. At higher temperatures the bipolaron becomes less mobile and pinned to a site. Without tweezers the bipolaron disappears quickly. (b) Total probability of the bipolaron summed across all sites, as a function of time. The total probability remains high when tweezers are on, indicating that the bipolaron is mobile without breaking up.}
    \label{evolution2d}
\end{figure*}

\emph{Experimental considerations}.---We consider a chain of ten \(^{40}\text{Ca}^+\) ions in a linear Paul trap, with secular trap frequencies \((\omega_x,\omega_y,\omega_z)=2\pi\times (3,4,0.5)\)~MHz. We encode the 4 possible spin states to the ion's long lived internal states as: $\ket{0}=\ket{^2D_{5/2},m_j=3/2}$, $\ket{\uparrow}=\ket{^2S_{1/2},m_j=1/2}$, $\ket{\downarrow}=\ket{^2S_{1/2},m_j=-1/2}$ and $\ket{\uparrow\downarrow}=\ket{^2D_{5/2},m_j=-3/2}$. 
We address each ion with an optical tweezer to modify the modes of the ion chain. A circularly-polarized tweezer with wavelength \(\lambda=630\)~nm, waist \(w_0=1\)~\(\mu\)m and power \(P_0=3\)~mW causes a tweezer trap frequency of \(\varpi=2\pi\times(7, 33, 32, -21\))~kHz on states \((\ket{0},\ket{\uparrow},\ket{\downarrow},\ket{\uparrow\downarrow}\)) respectively, where \((-)\) signifies anti-confinement. The off-resonant photon scattering remains below 1~s$^{-1}$. We apply site-dependent tweezer intensities to the ion chain with weights to make a near-homogeneous potential, as described in Fig. \ref{energies}. Setting the hopping term larger than the variation in site energy ensures spin mobility as soon as we add a hopping term. Specializing to the situation where there are two opposite spins in the crystal, we calculate the energy $B_{ij}$ for spin positions $i,j$ as shown in Fig.~\ref{energies}. Here, we assumed the groundstate of motion for the ion crystal and diagonalized the exact Hessian for each configuration.

For the hopping term $H_J$, we assume a radial trap frequency of $\omega_{\perp}=2\pi\times$3~MHz and we assume $\mu=2\pi\times$3.3~MHz, which results in $\alpha=1.43$. The Rabi frequency of the laser field generating the hopping is set as $\Omega_{\perp}=2\pi\times$150~kHz, which results in max($J_{ij})\sim$~20~Hz. 

We simulate the dynamics of Hamiltonian~(\ref{eq_Heff}) for a spin pair created at the 4$^\text{th}$ ion, i.e. this ion is prepared in state $\ket{\uparrow\downarrow}$ with all others in $\ket{0}$. The ion crystal is set to be in the ground state of motion. We calculate the probability of finding each ion in state $\ket{\uparrow\downarrow}$ as a function of time. The results are shown in Fig.~\ref{evolution2d}. We see that the bipolaron spreads through the crystal. In contrast, when we switch off the tweezer Hamiltonian the bipolaron quickly breaks up.

\emph{Bipolaron mobility}.---We now consider the case where the ion crystal is not prepared in the ground state of motion.  The hopping term is not affected to first order by thermal phonon occupation~\cite{Molmer:1999,Kirchmair:2009} and we therefore only consider the axial modes. 
For a crystal at temperature $T$, the average number of quanta in each mode is given by the Bose-Einstein distribution: $\bar{n}_m=1/(e^{\frac{\hbar\omega_m}{k_\text{B}T}}-1)$. The probability of finding the crystal in state $\mathbf{n}=(n_1,n_2,...,n_N)$ is given by: $P(\mathbf{n},T)=\prod_m^N \bar{n}_m^{n_m}/(1+\bar{n}_m)^{n_m+1}$.

\begin{figure}
    \centering
    \includegraphics[width=0.95\linewidth]{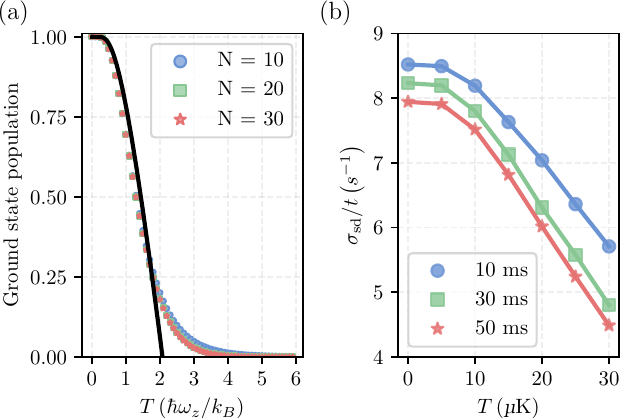}
    \caption{(a) Ground state population versus temperature in all modes except the center-of-mass mode for different ion cystal numbers \(N\), with the black line showing Eq. \ref{eq_Pmobile}. (b) Spread of the bipolaron \(\sigma_{sd}\) after waiting a fixed time \(t=10\)~ms, \(30\)~ms and \(50\)~ms, as a function of temperature.}
    \label{sd}
\end{figure}

As an example, we show the dynamics of a bipolaron in an ion crystal in a thermal state corresponding to $\bar{n}_\text{com}=$~0.43 in figure \ref{evolution2d}, where com stands for the center-of-mass mode with lowest frequency $\omega_z$. We see that in this case the bipolaron gets partially pinned to its starting location. This can be understood by studying  Hamiltonian~(\ref{eq_Hwithphonons}). The thermal occupation generates an inhomogeneous potential landscape over the ion crystal, causing the excitation to remain localized. The groundstate occupation and that of the homogeneous center-of-mass mode with $b_i=1/\sqrt{N}$ always contribute to the mobility. This may be quantified as: $P_\text{mobile}=\prod_{m=2}^N\left(1-e^{-\hbar\omega_m/(k_\text{B}T)}\right)$. In the low temperature limit, $P_\text{mobile}\approx 1-\sum_{m=2}^N e^{-\hbar\omega_m/k_\text{B}T}$. For long ion crystals, the mode frequencies can be approximated as $\omega_m\approx\omega_z m^\nu$~\cite{James:1998}. In the case of $\nu=1$, and in the limit $N\rightarrow \infty$, we can derive an expression for the bipolaron mobility:
\begin{equation}\label{eq_Pmobile}   
P_\text{mobile}\approx \frac{2\sinh{\left(\frac{\hbar\omega_z}{k_\text{B}T}\right)}-1}{e^{\frac{\hbar\omega_z}{k_\text{B}T}}-1}.
\end{equation} 
This shows that the bipolaron remain mobile at finite temperature $T\lesssim \hbar\omega_z/k_\text{B}$. For axial modes in a trapped ion crystal, we have $\nu \sim 0.81-0.83$ for $N=10-30$, but numerical analysis of the groundstate probability reveals that the temperature dependence is similar and almost independent of $N$ as shown in Fig.~\ref{sd}. We note that constellations of $\mathbf{n}$ besides the ground state may also contribute to mobility based on coincidence and the parameters of the hopping and tweezer Hamiltonian. 

We illustrate the mobility of the bipolaron in Fig. \ref{sd} by finding the standard deviation \(\sigma_\text{sd}=\sqrt{\frac{1}{N}\sum_{i=1}^N  (i-\bar{i})^2\mathcal{P}_i}\) in the spread of the bipolaron. Here, $\mathcal{P}_i$ denotes the probability to find the bipolaron at site $i$ and $\bar{i}$ is the initial site. The simulated range of temperatures lies below the Doppler limit of laser cooling of typical ions, $T_\text{D}\sim$~500~$\mu$K, but can be reached by sub-Doppler cooling techniques such as resolved sideband cooling~\cite{Wesenberg:1995}. Tuning of cooling parameters or controlled heating after ground state cooling can be employed to study the crossover between mobile and static bipolarons in the laboratory.

\emph{Conclusions}.---We have presented a proposal to simulate quadratic spin-phonon coupling in trapped ions by employing optical tweezers. In contrast to most trapped ion quantum simulators, the ion motion and the temperature of the ion crystal play a crucial role. To illustrate we have shown that a crossover between mobile and static bipolarons may be observed as a function of temperature. In this case, the zero-point energy of the ion crystal drives the emergence of mobile bipolarons. The system mimics the mechanism by which bipolarons emerge in solids due to quadratic electron-phonon coupling, predicted to result in higher $T_\text{c}$-values for superconductivity~\cite{Han:2024}. The presented setup may be further explored by considering multiple bipolarons and adding interactions between the simulated spins. Furthermore, on-site many-body interactions mediated by the quadratic spin-phonon coupling may be engineered by using more internal states.

\emph{Acknowledgements}.---We would like to acknowledge Ji\v{r}\'{i} Min\'{a}\v{r} and Jeremy T. Young for helpful feedback on the manuscript. This work was supported by the Netherlands Organization for Scientific Research (Grant Nos. 680.91.120, VI.C.202.051, OCENW.XL21.XL21.122 and No. OCENW.M.22.403). A.S.N is supported by the Dutch Research Council (NWO/OCW) as part of the Quantum Software Consortium programme (project number 024.003.037). A.S.N. is supported by Quantum Delta NL (project number NGF.1582.22.030).

\let\thefigureWithoutS\thefigure %% <- store old definition
\renewcommand\thefigure{A-\thefigureWithoutS}

%\renewcommand*{\citenumfont}[1]{A#1}
%\renewcommand*{\bibnumfmt}[1]{[A#1]}

%\section{Appendices}

\bibliographystyle{apsrev4-2}
\bibliography{biblio-RG}

\begin{thebibliography}{52}%
\makeatletter
\providecommand \@ifxundefined [1]{%
 \@ifx{#1\undefined}
}%
\providecommand \@ifnum [1]{%
 \ifnum #1\expandafter \@firstoftwo
 \else \expandafter \@secondoftwo
 \fi
}%
\providecommand \@ifx [1]{%
 \ifx #1\expandafter \@firstoftwo
 \else \expandafter \@secondoftwo
 \fi
}%
\providecommand \natexlab [1]{#1}%
\providecommand \enquote  [1]{``#1''}%
\providecommand \bibnamefont  [1]{#1}%
\providecommand \bibfnamefont [1]{#1}%
\providecommand \citenamefont [1]{#1}%
\providecommand \href@noop [0]{\@secondoftwo}%
\providecommand \href [0]{\begingroup \@sanitize@url \@href}%
\providecommand \@href[1]{\@@startlink{#1}\@@href}%
\providecommand \@@href[1]{\endgroup#1\@@endlink}%
\providecommand \@sanitize@url [0]{\catcode `\\12\catcode `\$12\catcode `\&12\catcode `\#12\catcode `\^12\catcode `\_12\catcode `\%12\relax}%
\providecommand \@@startlink[1]{}%
\providecommand \@@endlink[0]{}%
\providecommand \url  [0]{\begingroup\@sanitize@url \@url }%
\providecommand \@url [1]{\endgroup\@href {#1}{\urlprefix }}%
\providecommand \urlprefix  [0]{URL }%
\providecommand \Eprint [0]{\href }%
\providecommand \doibase [0]{https://doi.org/}%
\providecommand \selectlanguage [0]{\@gobble}%
\providecommand \bibinfo  [0]{\@secondoftwo}%
\providecommand \bibfield  [0]{\@secondoftwo}%
\providecommand \translation [1]{[#1]}%
\providecommand \BibitemOpen [0]{}%
\providecommand \bibitemStop [0]{}%
\providecommand \bibitemNoStop [0]{.\EOS\space}%
\providecommand \EOS [0]{\spacefactor3000\relax}%
\providecommand \BibitemShut  [1]{\csname bibitem#1\endcsname}%
\let\auto@bib@innerbib\@empty
%</preamble>
\bibitem [{\citenamefont {Porras}\ and\ \citenamefont {Cirac}(2004)}]{Porras_2004}%
  \BibitemOpen
  \bibfield  {author} {\bibinfo {author} {\bibfnamefont {D.}~\bibnamefont {Porras}}\ and\ \bibinfo {author} {\bibfnamefont {J.~I.}\ \bibnamefont {Cirac}},\ }\href {https://doi.org/10.1103/physrevlett.92.207901} {\bibfield  {journal} {\bibinfo  {journal} {Physical Review Letters}\ }\textbf {\bibinfo {volume} {92}},\ \bibinfo {pages} {207901} (\bibinfo {year} {2004})}\BibitemShut {NoStop}%
\bibitem [{\citenamefont {Friedenauer}\ \emph {et~al.}(2008)\citenamefont {Friedenauer}, \citenamefont {Schmitz}, \citenamefont {Glueckert}, \citenamefont {Porras},\ and\ \citenamefont {Schaetz}}]{Friedenauer:2008}%
  \BibitemOpen
  \bibfield  {author} {\bibinfo {author} {\bibfnamefont {H.}~\bibnamefont {Friedenauer}}, \bibinfo {author} {\bibfnamefont {H.}~\bibnamefont {Schmitz}}, \bibinfo {author} {\bibfnamefont {J.}~\bibnamefont {Glueckert}}, \bibinfo {author} {\bibfnamefont {D.}~\bibnamefont {Porras}},\ and\ \bibinfo {author} {\bibfnamefont {T.}~\bibnamefont {Schaetz}},\ }\href {https://doi.org/10.1038/nphys1032} {\bibfield  {journal} {\bibinfo  {journal} {Nat. Phys.}\ }\textbf {\bibinfo {volume} {4}},\ \bibinfo {pages} {757} (\bibinfo {year} {2008})}\BibitemShut {NoStop}%
\bibitem [{\citenamefont {Kim}\ \emph {et~al.}(2009)\citenamefont {Kim}, \citenamefont {Chang}, \citenamefont {Islam}, \citenamefont {Korenblit}, \citenamefont {Duan},\ and\ \citenamefont {Monroe}}]{Kim:2009}%
  \BibitemOpen
  \bibfield  {author} {\bibinfo {author} {\bibfnamefont {K.}~\bibnamefont {Kim}}, \bibinfo {author} {\bibfnamefont {M.-S.}\ \bibnamefont {Chang}}, \bibinfo {author} {\bibfnamefont {R.}~\bibnamefont {Islam}}, \bibinfo {author} {\bibfnamefont {S.}~\bibnamefont {Korenblit}}, \bibinfo {author} {\bibfnamefont {L.-M.}\ \bibnamefont {Duan}},\ and\ \bibinfo {author} {\bibfnamefont {C.}~\bibnamefont {Monroe}},\ }\href {https://doi.org/10.1103/PhysRevLett.103.120502} {\bibfield  {journal} {\bibinfo  {journal} {Phys.~Rev.~Lett.}\ }\textbf {\bibinfo {volume} {103}},\ \bibinfo {pages} {120502} (\bibinfo {year} {2009})}\BibitemShut {NoStop}%
\bibitem [{\citenamefont {Jurcevic}\ \emph {et~al.}(2014)\citenamefont {Jurcevic}, \citenamefont {Lanyon}, \citenamefont {Hauke}, \citenamefont {Hempel}, \citenamefont {Zoller}, \citenamefont {Blatt},\ and\ \citenamefont {Roos}}]{Jurcevic:2014}%
  \BibitemOpen
  \bibfield  {author} {\bibinfo {author} {\bibfnamefont {P.}~\bibnamefont {Jurcevic}}, \bibinfo {author} {\bibfnamefont {B.~P.}\ \bibnamefont {Lanyon}}, \bibinfo {author} {\bibfnamefont {P.}~\bibnamefont {Hauke}}, \bibinfo {author} {\bibfnamefont {C.}~\bibnamefont {Hempel}}, \bibinfo {author} {\bibfnamefont {P.}~\bibnamefont {Zoller}}, \bibinfo {author} {\bibfnamefont {R.}~\bibnamefont {Blatt}},\ and\ \bibinfo {author} {\bibfnamefont {C.~F.}\ \bibnamefont {Roos}},\ }\href {https://doi.org/10.1038/nature13461} {\bibfield  {journal} {\bibinfo  {journal} {Nature}\ }\textbf {\bibinfo {volume} {511}},\ \bibinfo {pages} {202} (\bibinfo {year} {2014})}\BibitemShut {NoStop}%
\bibitem [{\citenamefont {Richerme}\ \emph {et~al.}(2014)\citenamefont {Richerme}, \citenamefont {Gong}, \citenamefont {Lee}, \citenamefont {Senko}, \citenamefont {Smith}, \citenamefont {Foss-Feig}, \citenamefont {Michalakis}, \citenamefont {Gorshkov},\ and\ \citenamefont {Monroe}}]{Monroe_Correlations_2014}%
  \BibitemOpen
  \bibfield  {author} {\bibinfo {author} {\bibfnamefont {P.}~\bibnamefont {Richerme}}, \bibinfo {author} {\bibfnamefont {Z.-X.}\ \bibnamefont {Gong}}, \bibinfo {author} {\bibfnamefont {A.}~\bibnamefont {Lee}}, \bibinfo {author} {\bibfnamefont {C.}~\bibnamefont {Senko}}, \bibinfo {author} {\bibfnamefont {J.}~\bibnamefont {Smith}}, \bibinfo {author} {\bibfnamefont {M.}~\bibnamefont {Foss-Feig}}, \bibinfo {author} {\bibfnamefont {S.}~\bibnamefont {Michalakis}}, \bibinfo {author} {\bibfnamefont {A.~V.}\ \bibnamefont {Gorshkov}},\ and\ \bibinfo {author} {\bibfnamefont {C.}~\bibnamefont {Monroe}},\ }\href {https://www.nature.com/articles/nature13450} {\bibfield  {journal} {\bibinfo  {journal} {Nature}\ }\textbf {\bibinfo {volume} {511}},\ \bibinfo {pages} {198} (\bibinfo {year} {2014})}\BibitemShut {NoStop}%
\bibitem [{\citenamefont {Zhang}\ \emph {et~al.}(2017)\citenamefont {Zhang}, \citenamefont {Pagano}, \citenamefont {Hess}, \citenamefont {Kyprianidis}, \citenamefont {Becker}, \citenamefont {Kaplan}, \citenamefont {Gorshkov}, \citenamefont {Gong},\ and\ \citenamefont {Monroe}}]{Zhang:2017}%
  \BibitemOpen
  \bibfield  {author} {\bibinfo {author} {\bibfnamefont {J.}~\bibnamefont {Zhang}}, \bibinfo {author} {\bibfnamefont {G.}~\bibnamefont {Pagano}}, \bibinfo {author} {\bibfnamefont {P.~W.}\ \bibnamefont {Hess}}, \bibinfo {author} {\bibfnamefont {A.}~\bibnamefont {Kyprianidis}}, \bibinfo {author} {\bibfnamefont {P.}~\bibnamefont {Becker}}, \bibinfo {author} {\bibfnamefont {H.}~\bibnamefont {Kaplan}}, \bibinfo {author} {\bibfnamefont {A.~V.}\ \bibnamefont {Gorshkov}}, \bibinfo {author} {\bibfnamefont {Z.-X.}\ \bibnamefont {Gong}},\ and\ \bibinfo {author} {\bibfnamefont {C.}~\bibnamefont {Monroe}},\ }\href {https://doi.org/10.1038/nature24654} {\bibfield  {journal} {\bibinfo  {journal} {Nature}\ }\textbf {\bibinfo {volume} {551}},\ \bibinfo {pages} {601} (\bibinfo {year} {2017})}\BibitemShut {NoStop}%
\bibitem [{\citenamefont {Bohnet}\ \emph {et~al.}(2016)\citenamefont {Bohnet}, \citenamefont {Sawyer}, \citenamefont {Britton}, \citenamefont {Wall}, \citenamefont {Rey}, \citenamefont {Foss-Feig},\ and\ \citenamefont {Bollinger}}]{Bohnet:2016}%
  \BibitemOpen
  \bibfield  {author} {\bibinfo {author} {\bibfnamefont {J.~G.}\ \bibnamefont {Bohnet}}, \bibinfo {author} {\bibfnamefont {B.~C.}\ \bibnamefont {Sawyer}}, \bibinfo {author} {\bibfnamefont {J.~W.}\ \bibnamefont {Britton}}, \bibinfo {author} {\bibfnamefont {M.~L.}\ \bibnamefont {Wall}}, \bibinfo {author} {\bibfnamefont {A.~M.}\ \bibnamefont {Rey}}, \bibinfo {author} {\bibfnamefont {M.}~\bibnamefont {Foss-Feig}},\ and\ \bibinfo {author} {\bibfnamefont {J.~J.}\ \bibnamefont {Bollinger}},\ }\href {https://doi.org/10.1126/science.aad9958} {\bibfield  {journal} {\bibinfo  {journal} {Science}\ }\textbf {\bibinfo {volume} {352}},\ \bibinfo {pages} {1297–1301} (\bibinfo {year} {2016})}\BibitemShut {NoStop}%
\bibitem [{\citenamefont {Postler}\ \emph {et~al.}(2022)\citenamefont {Postler}, \citenamefont {Heu$\beta$en}, \citenamefont {Pogorelov}, \citenamefont {Rispler}, \citenamefont {Feldker}, \citenamefont {Meth}, \citenamefont {Marciniak}, \citenamefont {Stricker}, \citenamefont {Ringbauer}, \citenamefont {Blatt} \emph {et~al.}}]{Postler:2022}%
  \BibitemOpen
  \bibfield  {author} {\bibinfo {author} {\bibfnamefont {L.}~\bibnamefont {Postler}}, \bibinfo {author} {\bibfnamefont {S.}~\bibnamefont {Heu$\beta$en}}, \bibinfo {author} {\bibfnamefont {I.}~\bibnamefont {Pogorelov}}, \bibinfo {author} {\bibfnamefont {M.}~\bibnamefont {Rispler}}, \bibinfo {author} {\bibfnamefont {T.}~\bibnamefont {Feldker}}, \bibinfo {author} {\bibfnamefont {M.}~\bibnamefont {Meth}}, \bibinfo {author} {\bibfnamefont {C.~D.}\ \bibnamefont {Marciniak}}, \bibinfo {author} {\bibfnamefont {R.}~\bibnamefont {Stricker}}, \bibinfo {author} {\bibfnamefont {M.}~\bibnamefont {Ringbauer}}, \bibinfo {author} {\bibfnamefont {R.}~\bibnamefont {Blatt}}, \emph {et~al.},\ }\href {https://doi.org/10.1038/s41586-022-04721-1} {\bibfield  {journal} {\bibinfo  {journal} {Nature}\ }\textbf {\bibinfo {volume} {605}},\ \bibinfo {pages} {675} (\bibinfo {year} {2022})}\BibitemShut {NoStop}%
\bibitem [{\citenamefont {Postler}\ \emph {et~al.}(2024)\citenamefont {Postler}, \citenamefont {Butt}, \citenamefont {Pogorelov}, \citenamefont {Marciniak}, \citenamefont {Heu\ss{}en}, \citenamefont {Blatt}, \citenamefont {Schindler}, \citenamefont {Rispler}, \citenamefont {M\"uller},\ and\ \citenamefont {Monz}}]{Postler:2024}%
  \BibitemOpen
  \bibfield  {author} {\bibinfo {author} {\bibfnamefont {L.}~\bibnamefont {Postler}}, \bibinfo {author} {\bibfnamefont {F.}~\bibnamefont {Butt}}, \bibinfo {author} {\bibfnamefont {I.}~\bibnamefont {Pogorelov}}, \bibinfo {author} {\bibfnamefont {C.~D.}\ \bibnamefont {Marciniak}}, \bibinfo {author} {\bibfnamefont {S.}~\bibnamefont {Heu\ss{}en}}, \bibinfo {author} {\bibfnamefont {R.}~\bibnamefont {Blatt}}, \bibinfo {author} {\bibfnamefont {P.}~\bibnamefont {Schindler}}, \bibinfo {author} {\bibfnamefont {M.}~\bibnamefont {Rispler}}, \bibinfo {author} {\bibfnamefont {M.}~\bibnamefont {M\"uller}},\ and\ \bibinfo {author} {\bibfnamefont {T.}~\bibnamefont {Monz}},\ }\href {https://doi.org/10.1103/PRXQuantum.5.030326} {\bibfield  {journal} {\bibinfo  {journal} {PRX Quantum}\ }\textbf {\bibinfo {volume} {5}},\ \bibinfo {pages} {030326} (\bibinfo {year} {2024})}\BibitemShut {NoStop}%
\bibitem [{\citenamefont {Egan}\ \emph {et~al.}(2021)\citenamefont {Egan}, \citenamefont {Debroy}, \citenamefont {Noel}, \citenamefont {Risinger}, \citenamefont {Zhu}, \citenamefont {Biswas}, \citenamefont {Newman}, \citenamefont {Li}, \citenamefont {Brown}, \citenamefont {Cetina} \emph {et~al.}}]{Egan:2021}%
  \BibitemOpen
  \bibfield  {author} {\bibinfo {author} {\bibfnamefont {L.}~\bibnamefont {Egan}}, \bibinfo {author} {\bibfnamefont {D.~M.}\ \bibnamefont {Debroy}}, \bibinfo {author} {\bibfnamefont {C.}~\bibnamefont {Noel}}, \bibinfo {author} {\bibfnamefont {A.}~\bibnamefont {Risinger}}, \bibinfo {author} {\bibfnamefont {D.}~\bibnamefont {Zhu}}, \bibinfo {author} {\bibfnamefont {D.}~\bibnamefont {Biswas}}, \bibinfo {author} {\bibfnamefont {M.}~\bibnamefont {Newman}}, \bibinfo {author} {\bibfnamefont {M.}~\bibnamefont {Li}}, \bibinfo {author} {\bibfnamefont {K.~R.}\ \bibnamefont {Brown}}, \bibinfo {author} {\bibfnamefont {M.}~\bibnamefont {Cetina}}, \emph {et~al.},\ }\href {https://doi.org/10.1038/s41586-021-03928-y} {\bibfield  {journal} {\bibinfo  {journal} {Nature}\ }\textbf {\bibinfo {volume} {598}},\ \bibinfo {pages} {281} (\bibinfo {year} {2021})}\BibitemShut {NoStop}%
\bibitem [{\citenamefont {Pogorelov}\ \emph {et~al.}(2021)\citenamefont {Pogorelov}, \citenamefont {Feldker}, \citenamefont {Marciniak}, \citenamefont {Postler}, \citenamefont {Jacob}, \citenamefont {Krieglsteiner}, \citenamefont {Podlesnic}, \citenamefont {Meth}, \citenamefont {Negnevitsky}, \citenamefont {Stadler}, \citenamefont {H\"ofer}, \citenamefont {W\"achter}, \citenamefont {Lakhmanskiy}, \citenamefont {Blatt}, \citenamefont {Schindler},\ and\ \citenamefont {Monz}}]{Pogorelov:2021}%
  \BibitemOpen
  \bibfield  {author} {\bibinfo {author} {\bibfnamefont {I.}~\bibnamefont {Pogorelov}}, \bibinfo {author} {\bibfnamefont {T.}~\bibnamefont {Feldker}}, \bibinfo {author} {\bibfnamefont {C.~D.}\ \bibnamefont {Marciniak}}, \bibinfo {author} {\bibfnamefont {L.}~\bibnamefont {Postler}}, \bibinfo {author} {\bibfnamefont {G.}~\bibnamefont {Jacob}}, \bibinfo {author} {\bibfnamefont {O.}~\bibnamefont {Krieglsteiner}}, \bibinfo {author} {\bibfnamefont {V.}~\bibnamefont {Podlesnic}}, \bibinfo {author} {\bibfnamefont {M.}~\bibnamefont {Meth}}, \bibinfo {author} {\bibfnamefont {V.}~\bibnamefont {Negnevitsky}}, \bibinfo {author} {\bibfnamefont {M.}~\bibnamefont {Stadler}}, \bibinfo {author} {\bibfnamefont {B.}~\bibnamefont {H\"ofer}}, \bibinfo {author} {\bibfnamefont {C.}~\bibnamefont {W\"achter}}, \bibinfo {author} {\bibfnamefont {K.}~\bibnamefont {Lakhmanskiy}}, \bibinfo {author} {\bibfnamefont {R.}~\bibnamefont {Blatt}}, \bibinfo {author} {\bibfnamefont {P.}~\bibnamefont {Schindler}},\ and\ \bibinfo {author}
  {\bibfnamefont {T.}~\bibnamefont {Monz}},\ }\href {https://doi.org/10.1103/PRXQuantum.2.020343} {\bibfield  {journal} {\bibinfo  {journal} {PRX Quantum}\ }\textbf {\bibinfo {volume} {2}},\ \bibinfo {pages} {020343} (\bibinfo {year} {2021})}\BibitemShut {NoStop}%
\bibitem [{\citenamefont {Wright}\ \emph {et~al.}(2019)\citenamefont {Wright}, \citenamefont {Beck}, \citenamefont {Debnath}, \citenamefont {Amini}, \citenamefont {Nam}, \citenamefont {Grzesiak}, \citenamefont {Chen}, \citenamefont {Pisenti}, \citenamefont {Chmielewski}, \citenamefont {Collins} \emph {et~al.}}]{Wright:2019}%
  \BibitemOpen
  \bibfield  {author} {\bibinfo {author} {\bibfnamefont {K.}~\bibnamefont {Wright}}, \bibinfo {author} {\bibfnamefont {K.~M.}\ \bibnamefont {Beck}}, \bibinfo {author} {\bibfnamefont {S.}~\bibnamefont {Debnath}}, \bibinfo {author} {\bibfnamefont {J.}~\bibnamefont {Amini}}, \bibinfo {author} {\bibfnamefont {Y.}~\bibnamefont {Nam}}, \bibinfo {author} {\bibfnamefont {N.}~\bibnamefont {Grzesiak}}, \bibinfo {author} {\bibfnamefont {J.-S.}\ \bibnamefont {Chen}}, \bibinfo {author} {\bibfnamefont {N.}~\bibnamefont {Pisenti}}, \bibinfo {author} {\bibfnamefont {M.}~\bibnamefont {Chmielewski}}, \bibinfo {author} {\bibfnamefont {C.}~\bibnamefont {Collins}}, \emph {et~al.},\ }\href {https://doi.org/10.1038/s41467-019-13534-2} {\bibfield  {journal} {\bibinfo  {journal} {Nature communications}\ }\textbf {\bibinfo {volume} {10}},\ \bibinfo {pages} {5464} (\bibinfo {year} {2019})}\BibitemShut {NoStop}%
\bibitem [{\citenamefont {Gustiani}\ \emph {et~al.}(2024)\citenamefont {Gustiani}, \citenamefont {Leichtle}, \citenamefont {Mills}, \citenamefont {Miller}, \citenamefont {Grassie},\ and\ \citenamefont {Kashefi}}]{Gustiani:2024}%
  \BibitemOpen
  \bibfield  {author} {\bibinfo {author} {\bibfnamefont {C.}~\bibnamefont {Gustiani}}, \bibinfo {author} {\bibfnamefont {D.}~\bibnamefont {Leichtle}}, \bibinfo {author} {\bibfnamefont {D.}~\bibnamefont {Mills}}, \bibinfo {author} {\bibfnamefont {J.}~\bibnamefont {Miller}}, \bibinfo {author} {\bibfnamefont {R.}~\bibnamefont {Grassie}},\ and\ \bibinfo {author} {\bibfnamefont {E.}~\bibnamefont {Kashefi}},\ }\bibfield  {journal} {\bibinfo  {journal} {arXiv preprint arXiv:2410.24133}\ }\href {https://doi.org/10.48550/arXiv.2410.24133} {10.48550/arXiv.2410.24133} (\bibinfo {year} {2024})\BibitemShut {NoStop}%
\bibitem [{\citenamefont {Ballance}\ \emph {et~al.}(2016)\citenamefont {Ballance}, \citenamefont {Harty}, \citenamefont {Linke}, \citenamefont {Sepiol},\ and\ \citenamefont {Lucas}}]{Ballance:2016}%
  \BibitemOpen
  \bibfield  {author} {\bibinfo {author} {\bibfnamefont {C.~J.}\ \bibnamefont {Ballance}}, \bibinfo {author} {\bibfnamefont {T.~P.}\ \bibnamefont {Harty}}, \bibinfo {author} {\bibfnamefont {N.~M.}\ \bibnamefont {Linke}}, \bibinfo {author} {\bibfnamefont {M.~A.}\ \bibnamefont {Sepiol}},\ and\ \bibinfo {author} {\bibfnamefont {D.~M.}\ \bibnamefont {Lucas}},\ }\href {https://doi.org/10.1103/PhysRevLett.117.060504} {\bibfield  {journal} {\bibinfo  {journal} {Phys. Rev. Lett.}\ }\textbf {\bibinfo {volume} {117}},\ \bibinfo {pages} {060504} (\bibinfo {year} {2016})}\BibitemShut {NoStop}%
\bibitem [{\citenamefont {M{\o}lmer}\ and\ \citenamefont {S{\o}rensen}(1999)}]{Molmer:1999}%
  \BibitemOpen
  \bibfield  {author} {\bibinfo {author} {\bibfnamefont {K.}~\bibnamefont {M{\o}lmer}}\ and\ \bibinfo {author} {\bibfnamefont {A.}~\bibnamefont {S{\o}rensen}},\ }\href {https://doi.org/10.1103/PhysRevLett.82.1835} {\bibfield  {journal} {\bibinfo  {journal} {Phys.~Rev.~Lett.}\ }\textbf {\bibinfo {volume} {82}},\ \bibinfo {pages} {1835} (\bibinfo {year} {1999})}\BibitemShut {NoStop}%
\bibitem [{\citenamefont {Mazzanti}\ \emph {et~al.}(2021)\citenamefont {Mazzanti}, \citenamefont {Sch\"ussler}, \citenamefont {Arias~Espinoza}, \citenamefont {Wu}, \citenamefont {Gerritsma},\ and\ \citenamefont {Safavi-Naini}}]{Mazzanti:2021}%
  \BibitemOpen
  \bibfield  {author} {\bibinfo {author} {\bibfnamefont {M.}~\bibnamefont {Mazzanti}}, \bibinfo {author} {\bibfnamefont {R.~X.}\ \bibnamefont {Sch\"ussler}}, \bibinfo {author} {\bibfnamefont {J.~D.}\ \bibnamefont {Arias~Espinoza}}, \bibinfo {author} {\bibfnamefont {Z.}~\bibnamefont {Wu}}, \bibinfo {author} {\bibfnamefont {R.}~\bibnamefont {Gerritsma}},\ and\ \bibinfo {author} {\bibfnamefont {A.}~\bibnamefont {Safavi-Naini}},\ }\href {https://doi.org/10.1103/PhysRevLett.127.260502} {\bibfield  {journal} {\bibinfo  {journal} {Phys. Rev. Lett.}\ }\textbf {\bibinfo {volume} {127}},\ \bibinfo {pages} {260502} (\bibinfo {year} {2021})}\BibitemShut {NoStop}%
\bibitem [{\citenamefont {Katz}\ \emph {et~al.}(2022)\citenamefont {Katz}, \citenamefont {Cetina},\ and\ \citenamefont {Monroe}}]{Katz:2022}%
  \BibitemOpen
  \bibfield  {author} {\bibinfo {author} {\bibfnamefont {O.}~\bibnamefont {Katz}}, \bibinfo {author} {\bibfnamefont {M.}~\bibnamefont {Cetina}},\ and\ \bibinfo {author} {\bibfnamefont {C.}~\bibnamefont {Monroe}},\ }\href {https://doi.org/10.1103/PhysRevLett.129.063603} {\bibfield  {journal} {\bibinfo  {journal} {Phys. Rev. Lett.}\ }\textbf {\bibinfo {volume} {129}},\ \bibinfo {pages} {063603} (\bibinfo {year} {2022})}\BibitemShut {NoStop}%
\bibitem [{\citenamefont {Katz}\ \emph {et~al.}(2023)\citenamefont {Katz}, \citenamefont {Feng}, \citenamefont {Risinger}, \citenamefont {Monroe},\ and\ \citenamefont {Cetina}}]{Katz:2023}%
  \BibitemOpen
  \bibfield  {author} {\bibinfo {author} {\bibfnamefont {O.}~\bibnamefont {Katz}}, \bibinfo {author} {\bibfnamefont {L.}~\bibnamefont {Feng}}, \bibinfo {author} {\bibfnamefont {A.}~\bibnamefont {Risinger}}, \bibinfo {author} {\bibfnamefont {C.}~\bibnamefont {Monroe}},\ and\ \bibinfo {author} {\bibfnamefont {M.}~\bibnamefont {Cetina}},\ }\href {https://doi.org/10.1038/s41567-023-02102-7} {\bibfield  {journal} {\bibinfo  {journal} {Nature Physics}\ }\textbf {\bibinfo {volume} {19}},\ \bibinfo {pages} {1452–1458} (\bibinfo {year} {2023})}\BibitemShut {NoStop}%
\bibitem [{\citenamefont {Leibfried}\ \emph {et~al.}(2002)\citenamefont {Leibfried}, \citenamefont {DeMarco}, \citenamefont {Meyer}, \citenamefont {Rowe}, \citenamefont {Ben-Kish}, \citenamefont {Britton}, \citenamefont {Itano}, \citenamefont {Jelenkovi{\'c}}, \citenamefont {Langer}, \citenamefont {Rosenband},\ and\ \citenamefont {Wineland}}]{Leibfried:2002}%
  \BibitemOpen
  \bibfield  {author} {\bibinfo {author} {\bibfnamefont {D.}~\bibnamefont {Leibfried}}, \bibinfo {author} {\bibfnamefont {B.}~\bibnamefont {DeMarco}}, \bibinfo {author} {\bibfnamefont {V.}~\bibnamefont {Meyer}}, \bibinfo {author} {\bibfnamefont {M.}~\bibnamefont {Rowe}}, \bibinfo {author} {\bibfnamefont {A.}~\bibnamefont {Ben-Kish}}, \bibinfo {author} {\bibfnamefont {J.}~\bibnamefont {Britton}}, \bibinfo {author} {\bibfnamefont {W.~M.}\ \bibnamefont {Itano}}, \bibinfo {author} {\bibfnamefont {B.}~\bibnamefont {Jelenkovi{\'c}}}, \bibinfo {author} {\bibfnamefont {C.}~\bibnamefont {Langer}}, \bibinfo {author} {\bibfnamefont {T.}~\bibnamefont {Rosenband}},\ and\ \bibinfo {author} {\bibfnamefont {D.~J.}\ \bibnamefont {Wineland}},\ }\href@noop {} {\bibfield  {journal} {\bibinfo  {journal} {Phys.~Rev.~Lett.}\ }\textbf {\bibinfo {volume} {89}},\ \bibinfo {pages} {247901} (\bibinfo {year} {2002})}\BibitemShut {NoStop}%
\bibitem [{\citenamefont {Mezzacapo}\ \emph {et~al.}(2012)\citenamefont {Mezzacapo}, \citenamefont {Casanova}, \citenamefont {Lamata},\ and\ \citenamefont {Solano}}]{Mezzacapo:2012}%
  \BibitemOpen
  \bibfield  {author} {\bibinfo {author} {\bibfnamefont {A.}~\bibnamefont {Mezzacapo}}, \bibinfo {author} {\bibfnamefont {J.}~\bibnamefont {Casanova}}, \bibinfo {author} {\bibfnamefont {L.}~\bibnamefont {Lamata}},\ and\ \bibinfo {author} {\bibfnamefont {E.}~\bibnamefont {Solano}},\ }\href {https://doi.org/10.1103/PhysRevLett.109.200501} {\bibfield  {journal} {\bibinfo  {journal} {Phys. Rev. Lett.}\ }\textbf {\bibinfo {volume} {109}},\ \bibinfo {pages} {200501} (\bibinfo {year} {2012})}\BibitemShut {NoStop}%
\bibitem [{\citenamefont {Leibfried}\ \emph {et~al.}(2003)\citenamefont {Leibfried}, \citenamefont {Blatt}, \citenamefont {Monroe},\ and\ \citenamefont {Wineland}}]{Leibfried:2003}%
  \BibitemOpen
  \bibfield  {author} {\bibinfo {author} {\bibfnamefont {D.}~\bibnamefont {Leibfried}}, \bibinfo {author} {\bibfnamefont {R.}~\bibnamefont {Blatt}}, \bibinfo {author} {\bibfnamefont {C.}~\bibnamefont {Monroe}},\ and\ \bibinfo {author} {\bibfnamefont {D.}~\bibnamefont {Wineland}},\ }\href {https://doi.org/10.1103/RevModPhys.75.281} {\bibfield  {journal} {\bibinfo  {journal} {Rev. Mod. Phys.}\ }\textbf {\bibinfo {volume} {75}},\ \bibinfo {pages} {281} (\bibinfo {year} {2003})}\BibitemShut {NoStop}%
\bibitem [{\citenamefont {Grier}(2003)}]{Grier:2003}%
  \BibitemOpen
  \bibfield  {author} {\bibinfo {author} {\bibfnamefont {D.~G.}\ \bibnamefont {Grier}},\ }\href {https://doi.org/https://doi.org/10.1038/nature01935} {\bibfield  {journal} {\bibinfo  {journal} {Nature}\ }\textbf {\bibinfo {volume} {424}},\ \bibinfo {pages} {810} (\bibinfo {year} {2003})}\BibitemShut {NoStop}%
\bibitem [{\citenamefont {Scholl}\ \emph {et~al.}(2021)\citenamefont {Scholl}, \citenamefont {Schuler}, \citenamefont {Williams}, \citenamefont {Eberharter}, \citenamefont {Barredo}, \citenamefont {Schymik}, \citenamefont {Lienhard}, \citenamefont {Henry}, \citenamefont {Lang}, \citenamefont {Lahaye}, \citenamefont {L{\"a}uchli},\ and\ \citenamefont {Browaeys}}]{Scholl:2021}%
  \BibitemOpen
  \bibfield  {author} {\bibinfo {author} {\bibfnamefont {P.}~\bibnamefont {Scholl}}, \bibinfo {author} {\bibfnamefont {M.}~\bibnamefont {Schuler}}, \bibinfo {author} {\bibfnamefont {H.~J.}\ \bibnamefont {Williams}}, \bibinfo {author} {\bibfnamefont {A.~A.}\ \bibnamefont {Eberharter}}, \bibinfo {author} {\bibfnamefont {D.}~\bibnamefont {Barredo}}, \bibinfo {author} {\bibfnamefont {K.-N.}\ \bibnamefont {Schymik}}, \bibinfo {author} {\bibfnamefont {V.}~\bibnamefont {Lienhard}}, \bibinfo {author} {\bibfnamefont {L.-P.}\ \bibnamefont {Henry}}, \bibinfo {author} {\bibfnamefont {T.~C.}\ \bibnamefont {Lang}}, \bibinfo {author} {\bibfnamefont {T.}~\bibnamefont {Lahaye}}, \bibinfo {author} {\bibfnamefont {A.~M.}\ \bibnamefont {L{\"a}uchli}},\ and\ \bibinfo {author} {\bibfnamefont {A.}~\bibnamefont {Browaeys}},\ }\href {https://doi.org/10.1038/s41586-021-03585-1} {\bibfield  {journal} {\bibinfo  {journal} {Nature}\ }\textbf {\bibinfo {volume} {595}},\ \bibinfo {pages} {233} (\bibinfo {year} {2021})}\BibitemShut
  {NoStop}%
\bibitem [{\citenamefont {Bluvstein}\ \emph {et~al.}(2021)\citenamefont {Bluvstein}, \citenamefont {Omran}, \citenamefont {Levine}, \citenamefont {Keesling}, \citenamefont {Semeghini}, \citenamefont {Ebadi}, \citenamefont {Wang}, \citenamefont {Michailidis}, \citenamefont {Maskara}, \citenamefont {Ho}, \citenamefont {Choi}, \citenamefont {Serbyn}, \citenamefont {Greiner}, \citenamefont {Vuleti{\'c}},\ and\ \citenamefont {Lukin}}]{Bluvstein:2021}%
  \BibitemOpen
  \bibfield  {author} {\bibinfo {author} {\bibfnamefont {D.}~\bibnamefont {Bluvstein}}, \bibinfo {author} {\bibfnamefont {A.}~\bibnamefont {Omran}}, \bibinfo {author} {\bibfnamefont {H.}~\bibnamefont {Levine}}, \bibinfo {author} {\bibfnamefont {A.}~\bibnamefont {Keesling}}, \bibinfo {author} {\bibfnamefont {G.}~\bibnamefont {Semeghini}}, \bibinfo {author} {\bibfnamefont {S.}~\bibnamefont {Ebadi}}, \bibinfo {author} {\bibfnamefont {T.~T.}\ \bibnamefont {Wang}}, \bibinfo {author} {\bibfnamefont {A.~A.}\ \bibnamefont {Michailidis}}, \bibinfo {author} {\bibfnamefont {N.}~\bibnamefont {Maskara}}, \bibinfo {author} {\bibfnamefont {W.~W.}\ \bibnamefont {Ho}}, \bibinfo {author} {\bibfnamefont {S.}~\bibnamefont {Choi}}, \bibinfo {author} {\bibfnamefont {M.}~\bibnamefont {Serbyn}}, \bibinfo {author} {\bibfnamefont {M.}~\bibnamefont {Greiner}}, \bibinfo {author} {\bibfnamefont {V.}~\bibnamefont {Vuleti{\'c}}},\ and\ \bibinfo {author} {\bibfnamefont {M.~D.}\ \bibnamefont {Lukin}},\ }\href
  {https://doi.org/10.1126/science.abg2530} {\bibfield  {journal} {\bibinfo  {journal} {Science}\ }\textbf {\bibinfo {volume} {371}},\ \bibinfo {pages} {1355} (\bibinfo {year} {2021})}\BibitemShut {NoStop}%
\bibitem [{\citenamefont {Omran}\ \emph {et~al.}(2019)\citenamefont {Omran}, \citenamefont {Levine}, \citenamefont {Keesling}, \citenamefont {Semeghini}, \citenamefont {Wang}, \citenamefont {Ebadi}, \citenamefont {Bernien}, \citenamefont {Zibrov}, \citenamefont {Pichler}, \citenamefont {Choi}, \citenamefont {Cui}, \citenamefont {Rossignolo}, \citenamefont {Rembold}, \citenamefont {Montangero}, \citenamefont {Calarco}, \citenamefont {Endres}, \citenamefont {Greiner}, \citenamefont {Vuleti{\'c}},\ and\ \citenamefont {Lukin}}]{Omran:2019}%
  \BibitemOpen
  \bibfield  {author} {\bibinfo {author} {\bibfnamefont {A.}~\bibnamefont {Omran}}, \bibinfo {author} {\bibfnamefont {H.}~\bibnamefont {Levine}}, \bibinfo {author} {\bibfnamefont {A.}~\bibnamefont {Keesling}}, \bibinfo {author} {\bibfnamefont {G.}~\bibnamefont {Semeghini}}, \bibinfo {author} {\bibfnamefont {T.~T.}\ \bibnamefont {Wang}}, \bibinfo {author} {\bibfnamefont {S.}~\bibnamefont {Ebadi}}, \bibinfo {author} {\bibfnamefont {H.}~\bibnamefont {Bernien}}, \bibinfo {author} {\bibfnamefont {A.~S.}\ \bibnamefont {Zibrov}}, \bibinfo {author} {\bibfnamefont {H.}~\bibnamefont {Pichler}}, \bibinfo {author} {\bibfnamefont {S.}~\bibnamefont {Choi}}, \bibinfo {author} {\bibfnamefont {J.}~\bibnamefont {Cui}}, \bibinfo {author} {\bibfnamefont {M.}~\bibnamefont {Rossignolo}}, \bibinfo {author} {\bibfnamefont {P.}~\bibnamefont {Rembold}}, \bibinfo {author} {\bibfnamefont {S.}~\bibnamefont {Montangero}}, \bibinfo {author} {\bibfnamefont {T.}~\bibnamefont {Calarco}}, \bibinfo {author} {\bibfnamefont {M.}~\bibnamefont
  {Endres}}, \bibinfo {author} {\bibfnamefont {M.}~\bibnamefont {Greiner}}, \bibinfo {author} {\bibnamefont {Vuleti{\'c}}},\ and\ \bibinfo {author} {\bibfnamefont {M.~D.}\ \bibnamefont {Lukin}},\ }\href {https://doi.org/10.1126/science.aax9743} {\bibfield  {journal} {\bibinfo  {journal} {Science}\ }\textbf {\bibinfo {volume} {365}},\ \bibinfo {pages} {570} (\bibinfo {year} {2019})}\BibitemShut {NoStop}%
\bibitem [{\citenamefont {Ashkin}\ \emph {et~al.}(1986)\citenamefont {Ashkin}, \citenamefont {Dziedzic}, \citenamefont {Bjorkholm},\ and\ \citenamefont {Chu}}]{Ashkin:1986}%
  \BibitemOpen
  \bibfield  {author} {\bibinfo {author} {\bibfnamefont {A.}~\bibnamefont {Ashkin}}, \bibinfo {author} {\bibfnamefont {J.~M.}\ \bibnamefont {Dziedzic}}, \bibinfo {author} {\bibfnamefont {J.~E.}\ \bibnamefont {Bjorkholm}},\ and\ \bibinfo {author} {\bibfnamefont {S.}~\bibnamefont {Chu}},\ }\href {https://doi.org/10.1364/OL.11.000288} {\bibfield  {journal} {\bibinfo  {journal} {Optics letters}\ }\textbf {\bibinfo {volume} {11}},\ \bibinfo {pages} {288} (\bibinfo {year} {1986})}\BibitemShut {NoStop}%
\bibitem [{\citenamefont {Beugnon}\ \emph {et~al.}(2007)\citenamefont {Beugnon}, \citenamefont {Tuchendler}, \citenamefont {Marion}, \citenamefont {Ga{\"e}tan}, \citenamefont {Miroshnychenko}, \citenamefont {Sortais}, \citenamefont {Lance}, \citenamefont {Jones}, \citenamefont {Messin}, \citenamefont {Browaeys},\ and\ \citenamefont {Grangier}}]{Beugnon:2007}%
  \BibitemOpen
  \bibfield  {author} {\bibinfo {author} {\bibfnamefont {J.}~\bibnamefont {Beugnon}}, \bibinfo {author} {\bibfnamefont {C.}~\bibnamefont {Tuchendler}}, \bibinfo {author} {\bibfnamefont {H.}~\bibnamefont {Marion}}, \bibinfo {author} {\bibfnamefont {A.}~\bibnamefont {Ga{\"e}tan}}, \bibinfo {author} {\bibfnamefont {Y.}~\bibnamefont {Miroshnychenko}}, \bibinfo {author} {\bibfnamefont {Y.~R.}\ \bibnamefont {Sortais}}, \bibinfo {author} {\bibfnamefont {A.~M.}\ \bibnamefont {Lance}}, \bibinfo {author} {\bibfnamefont {M.~P.}\ \bibnamefont {Jones}}, \bibinfo {author} {\bibfnamefont {G.}~\bibnamefont {Messin}}, \bibinfo {author} {\bibfnamefont {A.}~\bibnamefont {Browaeys}},\ and\ \bibinfo {author} {\bibfnamefont {P.}~\bibnamefont {Grangier}},\ }\href {https://doi.org/10.1038/nphys698} {\bibfield  {journal} {\bibinfo  {journal} {Nature Physics}\ }\textbf {\bibinfo {volume} {3}},\ \bibinfo {pages} {696} (\bibinfo {year} {2007})}\BibitemShut {NoStop}%
\bibitem [{\citenamefont {Barredo}\ \emph {et~al.}(2016)\citenamefont {Barredo}, \citenamefont {De~L{\'e}s{\'e}leuc}, \citenamefont {Lienhard}, \citenamefont {Lahaye},\ and\ \citenamefont {Browaeys}}]{Barredo:2016}%
  \BibitemOpen
  \bibfield  {author} {\bibinfo {author} {\bibfnamefont {D.}~\bibnamefont {Barredo}}, \bibinfo {author} {\bibfnamefont {S.}~\bibnamefont {De~L{\'e}s{\'e}leuc}}, \bibinfo {author} {\bibfnamefont {V.}~\bibnamefont {Lienhard}}, \bibinfo {author} {\bibfnamefont {T.}~\bibnamefont {Lahaye}},\ and\ \bibinfo {author} {\bibfnamefont {A.}~\bibnamefont {Browaeys}},\ }\href {https://doi.org/https://doi.org/10.1126/science.aah3778} {\bibfield  {journal} {\bibinfo  {journal} {Science}\ }\textbf {\bibinfo {volume} {354}},\ \bibinfo {pages} {1021} (\bibinfo {year} {2016})}\BibitemShut {NoStop}%
\bibitem [{\citenamefont {Endres}\ \emph {et~al.}(2016)\citenamefont {Endres}, \citenamefont {Bernien}, \citenamefont {Keesling}, \citenamefont {Levine}, \citenamefont {Anschuetz}, \citenamefont {Krajenbrink}, \citenamefont {Senko}, \citenamefont {Vuleti{\'c}}, \citenamefont {Greiner},\ and\ \citenamefont {Lukin}}]{Endres:2016}%
  \BibitemOpen
  \bibfield  {author} {\bibinfo {author} {\bibfnamefont {M.}~\bibnamefont {Endres}}, \bibinfo {author} {\bibfnamefont {H.}~\bibnamefont {Bernien}}, \bibinfo {author} {\bibfnamefont {A.}~\bibnamefont {Keesling}}, \bibinfo {author} {\bibfnamefont {H.}~\bibnamefont {Levine}}, \bibinfo {author} {\bibfnamefont {E.~R.}\ \bibnamefont {Anschuetz}}, \bibinfo {author} {\bibfnamefont {A.}~\bibnamefont {Krajenbrink}}, \bibinfo {author} {\bibfnamefont {C.}~\bibnamefont {Senko}}, \bibinfo {author} {\bibfnamefont {V.}~\bibnamefont {Vuleti{\'c}}}, \bibinfo {author} {\bibfnamefont {M.}~\bibnamefont {Greiner}},\ and\ \bibinfo {author} {\bibfnamefont {M.~D.}\ \bibnamefont {Lukin}},\ }\href {https://doi.org/https://doi.org/10.1126/science.aah3752} {\bibfield  {journal} {\bibinfo  {journal} {Science}\ }\textbf {\bibinfo {volume} {354}},\ \bibinfo {pages} {1024} (\bibinfo {year} {2016})}\BibitemShut {NoStop}%
\bibitem [{\citenamefont {Graham}\ \emph {et~al.}(2019)\citenamefont {Graham}, \citenamefont {Kwon}, \citenamefont {Grinkemeyer}, \citenamefont {Marra}, \citenamefont {Jiang}, \citenamefont {Lichtman}, \citenamefont {Sun}, \citenamefont {Ebert},\ and\ \citenamefont {Saffman}}]{Graham:2019}%
  \BibitemOpen
  \bibfield  {author} {\bibinfo {author} {\bibfnamefont {T.~M.}\ \bibnamefont {Graham}}, \bibinfo {author} {\bibfnamefont {M.}~\bibnamefont {Kwon}}, \bibinfo {author} {\bibfnamefont {B.}~\bibnamefont {Grinkemeyer}}, \bibinfo {author} {\bibfnamefont {Z.}~\bibnamefont {Marra}}, \bibinfo {author} {\bibfnamefont {X.}~\bibnamefont {Jiang}}, \bibinfo {author} {\bibfnamefont {M.~T.}\ \bibnamefont {Lichtman}}, \bibinfo {author} {\bibfnamefont {Y.}~\bibnamefont {Sun}}, \bibinfo {author} {\bibfnamefont {M.}~\bibnamefont {Ebert}},\ and\ \bibinfo {author} {\bibfnamefont {M.}~\bibnamefont {Saffman}},\ }\href {https://doi.org/10.1103/PhysRevLett.123.230501} {\bibfield  {journal} {\bibinfo  {journal} {Phys. Rev. Lett.}\ }\textbf {\bibinfo {volume} {123}},\ \bibinfo {pages} {230501} (\bibinfo {year} {2019})}\BibitemShut {NoStop}%
\bibitem [{\citenamefont {Kaufman}\ and\ \citenamefont {Ni}(2021)}]{Kaufman:2021}%
  \BibitemOpen
  \bibfield  {author} {\bibinfo {author} {\bibfnamefont {A.~M.}\ \bibnamefont {Kaufman}}\ and\ \bibinfo {author} {\bibfnamefont {K.-K.}\ \bibnamefont {Ni}},\ }\href {https://doi.org/https://doi.org/10.1038/s41567-021-01357-2} {\bibfield  {journal} {\bibinfo  {journal} {Nature Physics}\ }\textbf {\bibinfo {volume} {17}},\ \bibinfo {pages} {1324} (\bibinfo {year} {2021})}\BibitemShut {NoStop}%
\bibitem [{\citenamefont {Stopp}\ \emph {et~al.}(2022)\citenamefont {Stopp}, \citenamefont {Verde}, \citenamefont {Katz}, \citenamefont {Drechsler}, \citenamefont {Schmiegelow},\ and\ \citenamefont {Schmidt-Kaler}}]{Stopp_ang_mom2022}%
  \BibitemOpen
  \bibfield  {author} {\bibinfo {author} {\bibfnamefont {F.}~\bibnamefont {Stopp}}, \bibinfo {author} {\bibfnamefont {M.}~\bibnamefont {Verde}}, \bibinfo {author} {\bibfnamefont {M.}~\bibnamefont {Katz}}, \bibinfo {author} {\bibfnamefont {M.}~\bibnamefont {Drechsler}}, \bibinfo {author} {\bibfnamefont {C.~T.}\ \bibnamefont {Schmiegelow}},\ and\ \bibinfo {author} {\bibfnamefont {F.}~\bibnamefont {Schmidt-Kaler}},\ }\href {https://doi.org/10.1103/PhysRevLett.129.263603} {\bibfield  {journal} {\bibinfo  {journal} {Phys. Rev. Lett.}\ }\textbf {\bibinfo {volume} {129}},\ \bibinfo {pages} {263603} (\bibinfo {year} {2022})}\BibitemShut {NoStop}%
\bibitem [{\citenamefont {Urech}\ \emph {et~al.}(2022)\citenamefont {Urech}, \citenamefont {Knottnerus}, \citenamefont {Spreeuw},\ and\ \citenamefont {Schreck}}]{Urech:2022}%
  \BibitemOpen
  \bibfield  {author} {\bibinfo {author} {\bibfnamefont {A.}~\bibnamefont {Urech}}, \bibinfo {author} {\bibfnamefont {I.~H.~A.}\ \bibnamefont {Knottnerus}}, \bibinfo {author} {\bibfnamefont {R.~J.~C.}\ \bibnamefont {Spreeuw}},\ and\ \bibinfo {author} {\bibfnamefont {F.}~\bibnamefont {Schreck}},\ }\href {https://doi.org/10.1103/PhysRevResearch.4.023245} {\bibfield  {journal} {\bibinfo  {journal} {Phys. Rev. Res.}\ }\textbf {\bibinfo {volume} {4}},\ \bibinfo {pages} {023245} (\bibinfo {year} {2022})}\BibitemShut {NoStop}%
\bibitem [{\citenamefont {luvstein}\ \emph {et~al.}(2024)\citenamefont {luvstein}, \citenamefont {Evered}, \citenamefont {Geim}, \citenamefont {Li}, \citenamefont {Zhou}, \citenamefont {Manovitz}, \citenamefont {Ebadi}, \citenamefont {Cain}, \citenamefont {Kalinowski}, \citenamefont {Hangleiter} \emph {et~al.}}]{bluvstein:2024}%
  \BibitemOpen
  \bibfield  {author} {\bibinfo {author} {\bibfnamefont {D.}~\bibnamefont {luvstein}}, \bibinfo {author} {\bibfnamefont {S.~J.}\ \bibnamefont {Evered}}, \bibinfo {author} {\bibfnamefont {A.~A.}\ \bibnamefont {Geim}}, \bibinfo {author} {\bibfnamefont {S.~H.}\ \bibnamefont {Li}}, \bibinfo {author} {\bibfnamefont {H.}~\bibnamefont {Zhou}}, \bibinfo {author} {\bibfnamefont {T.}~\bibnamefont {Manovitz}}, \bibinfo {author} {\bibfnamefont {S.}~\bibnamefont {Ebadi}}, \bibinfo {author} {\bibfnamefont {M.}~\bibnamefont {Cain}}, \bibinfo {author} {\bibfnamefont {M.}~\bibnamefont {Kalinowski}}, \bibinfo {author} {\bibfnamefont {D.}~\bibnamefont {Hangleiter}}, \emph {et~al.},\ }\href {https://doi.org/https://doi.org/10.1038/s41586-023-06927-3} {\bibfield  {journal} {\bibinfo  {journal} {Nature}\ }\textbf {\bibinfo {volume} {626}},\ \bibinfo {pages} {58} (\bibinfo {year} {2024})}\BibitemShut {NoStop}%
\bibitem [{\citenamefont {Teoh}\ \emph {et~al.}(2021)\citenamefont {Teoh}, \citenamefont {Sajjan}, \citenamefont {Sun}, \citenamefont {Rajabi},\ and\ \citenamefont {Islam}}]{Teoh:2021}%
  \BibitemOpen
  \bibfield  {author} {\bibinfo {author} {\bibfnamefont {Y.~H.}\ \bibnamefont {Teoh}}, \bibinfo {author} {\bibfnamefont {M.}~\bibnamefont {Sajjan}}, \bibinfo {author} {\bibfnamefont {Z.}~\bibnamefont {Sun}}, \bibinfo {author} {\bibfnamefont {F.}~\bibnamefont {Rajabi}},\ and\ \bibinfo {author} {\bibfnamefont {R.}~\bibnamefont {Islam}},\ }\href {https://doi.org/10.1103/PhysRevA.104.022420} {\bibfield  {journal} {\bibinfo  {journal} {Phys. Rev. A}\ }\textbf {\bibinfo {volume} {104}},\ \bibinfo {pages} {022420} (\bibinfo {year} {2021})}\BibitemShut {NoStop}%
\bibitem [{\citenamefont {{Arias Espinoza}}\ \emph {et~al.}(2021)\citenamefont {{Arias Espinoza}}, \citenamefont {Mazzanti}, \citenamefont {Fouka}, \citenamefont {Schüssler}, \citenamefont {Wu}, \citenamefont {Corboz}, \citenamefont {Gerritsma},\ and\ \citenamefont {Safavi-Naini}}]{Espinoza:2021}%
  \BibitemOpen
  \bibfield  {author} {\bibinfo {author} {\bibfnamefont {J.~D.}\ \bibnamefont {{Arias Espinoza}}}, \bibinfo {author} {\bibfnamefont {M.}~\bibnamefont {Mazzanti}}, \bibinfo {author} {\bibfnamefont {K.}~\bibnamefont {Fouka}}, \bibinfo {author} {\bibfnamefont {R.~X.}\ \bibnamefont {Schüssler}}, \bibinfo {author} {\bibfnamefont {Z.}~\bibnamefont {Wu}}, \bibinfo {author} {\bibfnamefont {P.}~\bibnamefont {Corboz}}, \bibinfo {author} {\bibfnamefont {R.}~\bibnamefont {Gerritsma}},\ and\ \bibinfo {author} {\bibfnamefont {A.}~\bibnamefont {Safavi-Naini}},\ }\href {https://doi.org/10.1103/PhysRevA.104.013302} {\bibfield  {journal} {\bibinfo  {journal} {Phys.~Rev.~A}\ }\textbf {\bibinfo {volume} {104}},\ \bibinfo {pages} {013302} (\bibinfo {year} {2021})}\BibitemShut {NoStop}%
\bibitem [{\citenamefont {Nath}\ \emph {et~al.}(2015)\citenamefont {Nath}, \citenamefont {Dalmonte}, \citenamefont {Glaetzle}, \citenamefont {Zoller}, \citenamefont {Schmidt-Kaler},\ and\ \citenamefont {Gerritsma}}]{Nath:2015}%
  \BibitemOpen
  \bibfield  {author} {\bibinfo {author} {\bibfnamefont {R.}~\bibnamefont {Nath}}, \bibinfo {author} {\bibfnamefont {M.}~\bibnamefont {Dalmonte}}, \bibinfo {author} {\bibfnamefont {A.~W.}\ \bibnamefont {Glaetzle}}, \bibinfo {author} {\bibfnamefont {P.}~\bibnamefont {Zoller}}, \bibinfo {author} {\bibfnamefont {F.}~\bibnamefont {Schmidt-Kaler}},\ and\ \bibinfo {author} {\bibfnamefont {R.}~\bibnamefont {Gerritsma}},\ }\href {https://doi.org/10.1088/1367-2630/17/6/065018} {\bibfield  {journal} {\bibinfo  {journal} {New Journal of Physics}\ }\textbf {\bibinfo {volume} {17}},\ \bibinfo {pages} {065018} (\bibinfo {year} {2015})}\BibitemShut {NoStop}%
\bibitem [{\citenamefont {Olsacher}\ \emph {et~al.}(2020)\citenamefont {Olsacher}, \citenamefont {Postler}, \citenamefont {Schindler}, \citenamefont {Monz}, \citenamefont {Zoller},\ and\ \citenamefont {Sieberer}}]{Olsacher:2020}%
  \BibitemOpen
  \bibfield  {author} {\bibinfo {author} {\bibfnamefont {T.}~\bibnamefont {Olsacher}}, \bibinfo {author} {\bibfnamefont {L.}~\bibnamefont {Postler}}, \bibinfo {author} {\bibfnamefont {P.}~\bibnamefont {Schindler}}, \bibinfo {author} {\bibfnamefont {T.}~\bibnamefont {Monz}}, \bibinfo {author} {\bibfnamefont {P.}~\bibnamefont {Zoller}},\ and\ \bibinfo {author} {\bibfnamefont {L.~M.}\ \bibnamefont {Sieberer}},\ }\href {https://doi.org/10.1103/PRXQuantum.1.020316} {\bibfield  {journal} {\bibinfo  {journal} {PRX Quantum}\ }\textbf {\bibinfo {volume} {1}},\ \bibinfo {pages} {020316} (\bibinfo {year} {2020})}\BibitemShut {NoStop}%
\bibitem [{\citenamefont {Bond}\ \emph {et~al.}(2022)\citenamefont {Bond}, \citenamefont {Lenstra}, \citenamefont {Gerritsma},\ and\ \citenamefont {Safavi-Naini}}]{Bond:2022}%
  \BibitemOpen
  \bibfield  {author} {\bibinfo {author} {\bibfnamefont {L.}~\bibnamefont {Bond}}, \bibinfo {author} {\bibfnamefont {L.}~\bibnamefont {Lenstra}}, \bibinfo {author} {\bibfnamefont {R.}~\bibnamefont {Gerritsma}},\ and\ \bibinfo {author} {\bibfnamefont {A.}~\bibnamefont {Safavi-Naini}},\ }\href {https://doi.org/10.1103/PhysRevA.106.042612} {\bibfield  {journal} {\bibinfo  {journal} {Phys. Rev. A}\ }\textbf {\bibinfo {volume} {106}},\ \bibinfo {pages} {042612} (\bibinfo {year} {2022})}\BibitemShut {NoStop}%
\bibitem [{\citenamefont {Schwerdt}\ \emph {et~al.}(2024)\citenamefont {Schwerdt}, \citenamefont {Peleg}, \citenamefont {Shapira}, \citenamefont {Priel}, \citenamefont {Florshaim}, \citenamefont {Gross}, \citenamefont {Zalic}, \citenamefont {Afek}, \citenamefont {Akerman}, \citenamefont {Stern}, \citenamefont {Kish},\ and\ \citenamefont {Ozeri}}]{Schwerdt:2024}%
  \BibitemOpen
  \bibfield  {author} {\bibinfo {author} {\bibfnamefont {D.}~\bibnamefont {Schwerdt}}, \bibinfo {author} {\bibfnamefont {L.}~\bibnamefont {Peleg}}, \bibinfo {author} {\bibfnamefont {Y.}~\bibnamefont {Shapira}}, \bibinfo {author} {\bibfnamefont {N.}~\bibnamefont {Priel}}, \bibinfo {author} {\bibfnamefont {Y.}~\bibnamefont {Florshaim}}, \bibinfo {author} {\bibfnamefont {A.}~\bibnamefont {Gross}}, \bibinfo {author} {\bibfnamefont {A.}~\bibnamefont {Zalic}}, \bibinfo {author} {\bibfnamefont {G.}~\bibnamefont {Afek}}, \bibinfo {author} {\bibfnamefont {N.}~\bibnamefont {Akerman}}, \bibinfo {author} {\bibfnamefont {A.}~\bibnamefont {Stern}}, \bibinfo {author} {\bibfnamefont {A.~B.}\ \bibnamefont {Kish}},\ and\ \bibinfo {author} {\bibfnamefont {R.}~\bibnamefont {Ozeri}},\ }\href {https://doi.org/10.1103/PhysRevX.14.041017} {\bibfield  {journal} {\bibinfo  {journal} {Phys. Rev. X}\ }\textbf {\bibinfo {volume} {14}},\ \bibinfo {pages} {041017} (\bibinfo {year} {2024})}\BibitemShut {NoStop}%
\bibitem [{\citenamefont {Han}\ \emph {et~al.}(2024)\citenamefont {Han}, \citenamefont {Kivelson},\ and\ \citenamefont {Volkov}}]{Han:2024}%
  \BibitemOpen
  \bibfield  {author} {\bibinfo {author} {\bibfnamefont {Z.}~\bibnamefont {Han}}, \bibinfo {author} {\bibfnamefont {S.~A.}\ \bibnamefont {Kivelson}},\ and\ \bibinfo {author} {\bibfnamefont {P.~A.}\ \bibnamefont {Volkov}},\ }\href {https://doi.org/10.1103/PhysRevLett.132.226001} {\bibfield  {journal} {\bibinfo  {journal} {Phys. Rev. Lett.}\ }\textbf {\bibinfo {volume} {132}},\ \bibinfo {pages} {226001} (\bibinfo {year} {2024})}\BibitemShut {NoStop}%
\bibitem [{\citenamefont {Zhang}\ \emph {et~al.}(2023)\citenamefont {Zhang}, \citenamefont {Kuklov}, \citenamefont {Prokof'ev},\ and\ \citenamefont {Svistunov}}]{Zhang:2023}%
  \BibitemOpen
  \bibfield  {author} {\bibinfo {author} {\bibfnamefont {Z.}~\bibnamefont {Zhang}}, \bibinfo {author} {\bibfnamefont {A.}~\bibnamefont {Kuklov}}, \bibinfo {author} {\bibfnamefont {N.}~\bibnamefont {Prokof'ev}},\ and\ \bibinfo {author} {\bibfnamefont {B.}~\bibnamefont {Svistunov}},\ }\href {https://doi.org/10.1103/PhysRevB.108.245127} {\bibfield  {journal} {\bibinfo  {journal} {Phys. Rev. B}\ }\textbf {\bibinfo {volume} {108}},\ \bibinfo {pages} {245127} (\bibinfo {year} {2023})}\BibitemShut {NoStop}%
\bibitem [{\citenamefont {Zhang}\ \emph {et~al.}(2024)\citenamefont {Zhang}, \citenamefont {Kuklov}, \citenamefont {Prokof'ev},\ and\ \citenamefont {Svistunov}}]{Zhang:2024}%
  \BibitemOpen
  \bibfield  {author} {\bibinfo {author} {\bibfnamefont {Z.}~\bibnamefont {Zhang}}, \bibinfo {author} {\bibfnamefont {A.}~\bibnamefont {Kuklov}}, \bibinfo {author} {\bibfnamefont {N.}~\bibnamefont {Prokof'ev}},\ and\ \bibinfo {author} {\bibfnamefont {B.}~\bibnamefont {Svistunov}},\ }\bibfield  {journal} {\bibinfo  {journal} {arXiv preprint arXiv:2408.03266}\ }\href {https://doi.org/10.48550/arXiv.2408.03266} {10.48550/arXiv.2408.03266} (\bibinfo {year} {2024})\BibitemShut {NoStop}%
\bibitem [{\citenamefont {Ragni}\ \emph {et~al.}(2023)\citenamefont {Ragni}, \citenamefont {Hahn}, \citenamefont {Zhang}, \citenamefont {Prokof'ev}, \citenamefont {Kuklov}, \citenamefont {Klimin}, \citenamefont {Houtput}, \citenamefont {Svistunov}, \citenamefont {Tempere}, \citenamefont {Nagaosa}, \citenamefont {Franchini},\ and\ \citenamefont {Mishchenko}}]{Ragni:2023}%
  \BibitemOpen
  \bibfield  {author} {\bibinfo {author} {\bibfnamefont {S.}~\bibnamefont {Ragni}}, \bibinfo {author} {\bibfnamefont {T.}~\bibnamefont {Hahn}}, \bibinfo {author} {\bibfnamefont {Z.}~\bibnamefont {Zhang}}, \bibinfo {author} {\bibfnamefont {N.}~\bibnamefont {Prokof'ev}}, \bibinfo {author} {\bibfnamefont {A.}~\bibnamefont {Kuklov}}, \bibinfo {author} {\bibfnamefont {S.}~\bibnamefont {Klimin}}, \bibinfo {author} {\bibfnamefont {M.}~\bibnamefont {Houtput}}, \bibinfo {author} {\bibfnamefont {B.}~\bibnamefont {Svistunov}}, \bibinfo {author} {\bibfnamefont {J.}~\bibnamefont {Tempere}}, \bibinfo {author} {\bibfnamefont {N.}~\bibnamefont {Nagaosa}}, \bibinfo {author} {\bibfnamefont {C.}~\bibnamefont {Franchini}},\ and\ \bibinfo {author} {\bibfnamefont {A.~S.}\ \bibnamefont {Mishchenko}},\ }\href {https://doi.org/10.1103/PhysRevB.107.L121109} {\bibfield  {journal} {\bibinfo  {journal} {Phys. Rev. B}\ }\textbf {\bibinfo {volume} {107}},\ \bibinfo {pages} {L121109} (\bibinfo {year} {2023})}\BibitemShut {NoStop}%
\bibitem [{\citenamefont {Smith}\ \emph {et~al.}(2016)\citenamefont {Smith}, \citenamefont {Lee}, \citenamefont {Richerme}, \citenamefont {Neyenhuis}, \citenamefont {Hess}, \citenamefont {Hauke}, \citenamefont {Heyl}, \citenamefont {Huse},\ and\ \citenamefont {Monroe}}]{Smith:2016}%
  \BibitemOpen
  \bibfield  {author} {\bibinfo {author} {\bibfnamefont {J.}~\bibnamefont {Smith}}, \bibinfo {author} {\bibfnamefont {A.}~\bibnamefont {Lee}}, \bibinfo {author} {\bibfnamefont {P.}~\bibnamefont {Richerme}}, \bibinfo {author} {\bibfnamefont {B.}~\bibnamefont {Neyenhuis}}, \bibinfo {author} {\bibfnamefont {P.~W.}\ \bibnamefont {Hess}}, \bibinfo {author} {\bibfnamefont {P.}~\bibnamefont {Hauke}}, \bibinfo {author} {\bibfnamefont {M.}~\bibnamefont {Heyl}}, \bibinfo {author} {\bibfnamefont {D.~A.}\ \bibnamefont {Huse}},\ and\ \bibinfo {author} {\bibfnamefont {C.}~\bibnamefont {Monroe}},\ }\href {https://doi.org/10.1038/nphys3783} {\bibfield  {journal} {\bibinfo  {journal} {Nat. Phys.}\ }\textbf {\bibinfo {volume} {12}},\ \bibinfo {pages} {907–911} (\bibinfo {year} {2016})}\BibitemShut {NoStop}%
\bibitem [{\citenamefont {Morong}\ \emph {et~al.}(2021)\citenamefont {Morong}, \citenamefont {Liu}, \citenamefont {Becker}, \citenamefont {Collins}, \citenamefont {Feng}, \citenamefont {Kyprianidis}, \citenamefont {Pagano}, \citenamefont {You}, \citenamefont {Gorshkov},\ and\ \citenamefont {Monroe}}]{Morong:2024}%
  \BibitemOpen
  \bibfield  {author} {\bibinfo {author} {\bibfnamefont {W.}~\bibnamefont {Morong}}, \bibinfo {author} {\bibfnamefont {F.}~\bibnamefont {Liu}}, \bibinfo {author} {\bibfnamefont {P.}~\bibnamefont {Becker}}, \bibinfo {author} {\bibfnamefont {K.~S.}\ \bibnamefont {Collins}}, \bibinfo {author} {\bibfnamefont {L.}~\bibnamefont {Feng}}, \bibinfo {author} {\bibfnamefont {A.}~\bibnamefont {Kyprianidis}}, \bibinfo {author} {\bibfnamefont {G.}~\bibnamefont {Pagano}}, \bibinfo {author} {\bibfnamefont {T.}~\bibnamefont {You}}, \bibinfo {author} {\bibfnamefont {A.~V.}\ \bibnamefont {Gorshkov}},\ and\ \bibinfo {author} {\bibfnamefont {C.}~\bibnamefont {Monroe}},\ }\href {https://doi.org/10.1038/s41586-021-03988-0} {\bibfield  {journal} {\bibinfo  {journal} {Nature}\ }\textbf {\bibinfo {volume} {599}},\ \bibinfo {pages} {393–398} (\bibinfo {year} {2021})}\BibitemShut {NoStop}%
\bibitem [{\citenamefont {Mazzanti}\ \emph {et~al.}(2023)\citenamefont {Mazzanti}, \citenamefont {Gerritsma}, \citenamefont {Spreeuw},\ and\ \citenamefont {Safavi-Naini}}]{Mazzanti:2023}%
  \BibitemOpen
  \bibfield  {author} {\bibinfo {author} {\bibfnamefont {M.}~\bibnamefont {Mazzanti}}, \bibinfo {author} {\bibfnamefont {R.}~\bibnamefont {Gerritsma}}, \bibinfo {author} {\bibfnamefont {R.~J.~C.}\ \bibnamefont {Spreeuw}},\ and\ \bibinfo {author} {\bibfnamefont {A.}~\bibnamefont {Safavi-Naini}},\ }\href {https://doi.org/10.1103/PhysRevResearch.5.033036} {\bibfield  {journal} {\bibinfo  {journal} {Phys. Rev. Res.}\ }\textbf {\bibinfo {volume} {5}},\ \bibinfo {pages} {033036} (\bibinfo {year} {2023})}\BibitemShut {NoStop}%
\bibitem [{\citenamefont {James}(1998)}]{James:1998}%
  \BibitemOpen
  \bibfield  {author} {\bibinfo {author} {\bibfnamefont {D.~F.~V.}\ \bibnamefont {James}},\ }\href {https://doi.org/10.1007/s003400050373} {\bibfield  {journal} {\bibinfo  {journal} {Appl.~Phys.~B}\ }\textbf {\bibinfo {volume} {66}},\ \bibinfo {pages} {181} (\bibinfo {year} {1998})}\BibitemShut {NoStop}%
\bibitem [{\citenamefont {Grimm}\ \emph {et~al.}(2000)\citenamefont {Grimm}, \citenamefont {Weidem{\"u}ller},\ and\ \citenamefont {Ovchinnikov}}]{Grimm:2000}%
  \BibitemOpen
  \bibfield  {author} {\bibinfo {author} {\bibfnamefont {R.}~\bibnamefont {Grimm}}, \bibinfo {author} {\bibfnamefont {M.}~\bibnamefont {Weidem{\"u}ller}},\ and\ \bibinfo {author} {\bibfnamefont {Y.~B.}\ \bibnamefont {Ovchinnikov}},\ }\href {https://doi.org/10.1016/S1049-250X(08)60186-X} {\bibfield  {journal} {\bibinfo  {journal} {Adv. At., Mol., Opt. Phys.}\ }\textbf {\bibinfo {volume} {42}},\ \bibinfo {pages} {95} (\bibinfo {year} {2000})}\BibitemShut {NoStop}%
\bibitem [{\citenamefont {Britton}\ \emph {et~al.}(2012)\citenamefont {Britton}, \citenamefont {Sawyer}, \citenamefont {A.~C.~Keith}, \citenamefont {Freericks}, \citenamefont {Uys}, \citenamefont {Biercuk},\ and\ \citenamefont {Bollinger}}]{Britton:2012}%
  \BibitemOpen
  \bibfield  {author} {\bibinfo {author} {\bibfnamefont {J.}~\bibnamefont {Britton}}, \bibinfo {author} {\bibfnamefont {B.~C.}\ \bibnamefont {Sawyer}}, \bibinfo {author} {\bibfnamefont {C.-C.~J.~W.}\ \bibnamefont {A.~C.~Keith}}, \bibinfo {author} {\bibfnamefont {J.}~\bibnamefont {Freericks}}, \bibinfo {author} {\bibfnamefont {H.}~\bibnamefont {Uys}}, \bibinfo {author} {\bibfnamefont {M.~J.}\ \bibnamefont {Biercuk}},\ and\ \bibinfo {author} {\bibfnamefont {J.~J.}\ \bibnamefont {Bollinger}},\ }\href {https://doi.org/10.1038/nature10981} {\bibfield  {journal} {\bibinfo  {journal} {Nature}\ }\textbf {\bibinfo {volume} {484}},\ \bibinfo {pages} {489} (\bibinfo {year} {2012})}\BibitemShut {NoStop}%
\bibitem [{\citenamefont {Kirchmair}\ \emph {et~al.}(2009)\citenamefont {Kirchmair}, \citenamefont {Benhelm}, \citenamefont {Z\"ahringer}, \citenamefont {Gerritsma}, \citenamefont {Roos},\ and\ \citenamefont {Blatt}}]{Kirchmair:2009}%
  \BibitemOpen
  \bibfield  {author} {\bibinfo {author} {\bibfnamefont {G.}~\bibnamefont {Kirchmair}}, \bibinfo {author} {\bibfnamefont {J.}~\bibnamefont {Benhelm}}, \bibinfo {author} {\bibfnamefont {F.}~\bibnamefont {Z\"ahringer}}, \bibinfo {author} {\bibfnamefont {R.}~\bibnamefont {Gerritsma}}, \bibinfo {author} {\bibfnamefont {C.~F.}\ \bibnamefont {Roos}},\ and\ \bibinfo {author} {\bibfnamefont {R.}~\bibnamefont {Blatt}},\ }\href {https://doi.org/10.1088/1367-2630/11/2/023002} {\bibfield  {journal} {\bibinfo  {journal} {New J. Phys.}\ }\textbf {\bibinfo {volume} {11}},\ \bibinfo {pages} {023002} (\bibinfo {year} {2009})}\BibitemShut {NoStop}%
\bibitem [{\citenamefont {Itano}\ \emph {et~al.}(1995)\citenamefont {Itano}, \citenamefont {Bergquist}, \citenamefont {Bollinger},\ and\ \citenamefont {Wineland}}]{Wesenberg:1995}%
  \BibitemOpen
  \bibfield  {author} {\bibinfo {author} {\bibfnamefont {W.~M.}\ \bibnamefont {Itano}}, \bibinfo {author} {\bibfnamefont {J.~C.}\ \bibnamefont {Bergquist}}, \bibinfo {author} {\bibfnamefont {J.~J.}\ \bibnamefont {Bollinger}},\ and\ \bibinfo {author} {\bibfnamefont {D.~J.}\ \bibnamefont {Wineland}},\ }\href {https://doi.org/10.1088/0031-8949/1995/T59/013} {\bibfield  {journal} {\bibinfo  {journal} {Phys. Scr.}\ }\textbf {\bibinfo {volume} {T59}},\ \bibinfo {pages} {106} (\bibinfo {year} {1995})}\BibitemShut {NoStop}%
\end{thebibliography}%

\end{document}